\documentclass[a4paper,fleqn]{cas-dc}

\usepackage[numbers]{natbib}

\def\tsc#1{\csdef{#1}{\textsc{\lowercase{#1}}\xspace}}
\tsc{WGM}
\tsc{QE}
\tsc{EP}
\tsc{PMS}
\tsc{BEC}
\tsc{DE}

\begin{document}
\let\WriteBookmarks\relax
\def\floatpagepagefraction{1}
\def\textpagefraction{.001}
\shorttitle{In-Vehicle Digital Twin-Based Collision Warning Framework with Sybil Attack Detection}
\shortauthors{Hasan et~al.}

\title [mode = title]{In-Vehicle Digital Twin-Based Collision Warning Framework with Sybil Attack Detection }                      


\author[1]{Mohammad Imtiaz Hasan}[orcid=0009-0006-9799-6773]
\cormark[1]
\ead{hasan2@clemson.edu}

\credit{Conceptualization, Methodology, Software, Validation, Formal analysis, Investigation, Data curation, Writing-original draft, Writing-review and editing, Visualization, Supervision}

\affiliation[1]{organization={Glenn Department of Civil Engineering, Clemson University},
                city={Clemson},
                postcode={29634}, 
                state={South Carolina},
                country={USA}}
                
\author[1]{Abyad Enan}[orcid=0000-0002-1599-5472]
\ead{aenan@clemson.edu}
\credit{ Validation, Investigation, Writing-review and editing, Visualization}

\author[1]{Jean Michel Tine}[orcid=0009-0001-1121-7323]
\ead{jtine@clemson.edu}
\credit{Validation, Investigation, Writing-review and editing, Visualization}

\author[1]{Araf Rahman}[orcid=0009-0006-2550-8208]
\ead{arafr@clemson.edu}
\credit{Validation, Investigation, Writing-review and editing, Visualization}

\author[1]{M Sabbir Salek}[orcid=0000-0001-7326-3694]
\ead{msalek@clemson.edu}
\credit{Conceptualization, Methodology, Validation, Investigation, Writing-review and editing, Visualization, Supervision}

\author[1]{Mashrur Chowdhury}[orcid=0000-0002-3275-6983]
\ead{mac@clemson.edu}
\credit{Conceptualization, Methodology, Validation, Investigation, Writing-review and editing, Visualization, Supervision}

\cortext[cor1]{Corresponding author}

\begin{abstract}
Connected Vehicles (CVs) rely extensively on communication technologies to enable data-driven predictive analyses for enhancing performance and safety. These communication channels can be exploited by adversaries to launch cyberattacks such as Sybil attacks, which could threaten both safety-critical and mobility applications, leaving CVs vulnerable and putting human lives at risk. As CV deployment continues to expand, the need to detect and mitigate cyberattacks in real-time becomes increasingly urgent. This study presents an in-vehicle Digital Twin (DT)-based collision warning framework with built-in capabilities for Sybil attacks detection. The framework integrates a Temporal Convolutional Network (TCN) for learning temporal dependencies in vehicle trajectory data and Hierarchical Navigable Small World (HNSW) algorithms for efficient similarity-based classification. Our framework is evaluated on real-world Sybil attack data, collected through field experiments. The framework achieved accuracy, recall, and F1 scores of 0.984, 1.00, and 0.944, respectively, in detecting Sybil-generated fake vehicles. During the safety evaluation, the framework reduced the mean Time Exposed Time-To-Collision (TET) and mean Time Integrated Time-To-Collision (TIT) of near-collision events by 88\% and 72\%, respectively. Furthermore, real-world feasibility evaluation shows that the framework conformed to the standardized maximum allowable latency for safety applications and operated well within the capacity of modern processors, demonstrating the promise of an in-vehicle DT-based framework as an attack-mitigation mechanism against Sybil attacks for next-generation CVs. 

\end{abstract}


\begin{highlights}
\item Developed a DT-based, fully in-vehicle framework that jointly detects and mitigates Sybil attacks while enabling collision risk assessment
\item Developed a real-time, infrastructure-independent Sybil detection approach that leverages spatio-temporal vehicle trajectory modeling
\item Validated the framework through field experiments under static and dynamic Sybil attack scenarios
\end{highlights}


\maketitle

\section{Introduction}
Connected vehicle (CV) communication technologies, such as Cellular Vehicle-to-Everything (C-V2X), serve as a fundamental component of intelligent transportation systems (ITS). These technologies enable vehicles to exchange real-time kinematic information, including position, speed, and heading direction, with other vehicles and roadside infrastructure, promising significant improvements in road safety and operational efficiency \ \cite{1}. Applications such as cooperative adaptive cruise control and intelligent traffic and intersection management systems leverage CV communication to enhance safety and mobility. In particular, C-V2X technology enables vehicles to both directly communicate over short ranges and relay information via cellular networks, creating an interconnected and complex ecosystem \cite{2}.

A key safety feature in modern vehicles is the collision warning application, which alerts drivers and/or autonomously initiates maneuvers to prevent accidents. These systems rely on real-time information to predict and avoid collisions. Traditional collision warning systems primarily rely on on-board perception sensors, such as radar, lidar, and cameras, to estimate inter-vehicle distance and collision risk. However, their performance can degrade under adverse weather conditions, occlusions, limited sensing range or limited line-of-sight situations. In contrast, C-V2X-based collision warning application is not affected by weather conditions, line-of-sight or occlusion and has a longer range compared to onboard sensors \cite{3}. 

\begin{figure}
	\centering
	\includegraphics[width=\columnwidth]{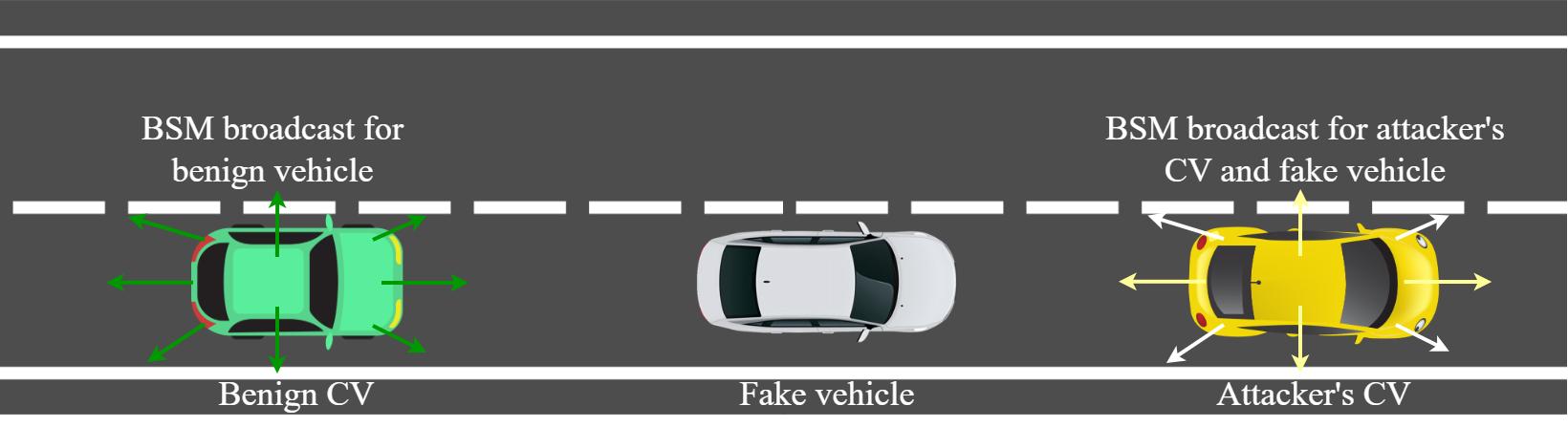}
	\caption{Sybil attack scenario.}
	\label{FIG1}
\end{figure}

However, C-V2X-enabled systems also inherit the vulnerabilities of networked environments, exposing them to a range of cyberattacks, such as Denial-of-Service (DoS), remote code execution (RCE), and Sybil attacks \cite{4,5}. Malicious actors can conduct these attacks to inject false information, disrupt data flows, disable safety-critical functions, or even take control of the vehicle. A particularly threatening scenario is the Sybil attack \cite{6}, where an attacker generates multiple fake vehicle identities and transmits forged Basic Safety Messages (BSMs). BSMs are standardized, frequently broadcast digital messages that share essential data, such as speed, location, and heading direction, using a C-V2X radio \cite{7}. In a Sybil attack, the adversary injects several counterfeit BSM streams that appear to originate from distinct vehicles, even though they are all controlled by the same physical entity. Fig. \ref{FIG1} illustrates a Sybil attack scenario in which the attacker CV is broadcasting BSM with its own ID and a fake vehicle ID. The benign vehicle is receiving BSMs and perceives two separate vehicles ahead, whereas in reality, there is only one. Such artificial inflation of perceived traffic can cause surrounding CVs and infrastructure to misinterpret the traffic situation, severely compromising the performance of safety-critical applications in a CV environment \cite{8}. In a C-V2X-based collision warning system, such attacks can result in false collision alerts or unnecessary evasive maneuvers such as sudden braking or lane changes, leading to unsafe road conditions or secondary collisions. Since Sybil attacks can have severe implications for traffic safety and security, an effective and robust Sybil detection approach is of paramount importance in a CV environment \cite{9}. Most of the current research on Sybil attack detection in vehicular environments relies on stationary infrastructure, such as roadside units (RSUs) or centralized servers (CSs), to manage authentication tokens, monitor location discrepancies, challenge suspected Sybil nodes, or run machine learning (ML)-based detection algorithms \cite{10,11,12}. However, the performance of these infrastructure-based methods can be significantly degraded by real-world obstacles, such as buildings and trees, which cause signal attenuation and communication loss around RSUs \cite{13}. In addition, RSUs are susceptible to physical and cyberattacks, which can create a single point of failure \cite{14}. These challenges can lead to unreliable detection performance in obstructed areas, underscoring the need for an infrastructure-independent Sybil detection approach.

\begin{figure}
	\centering
	\includegraphics[width=\columnwidth]{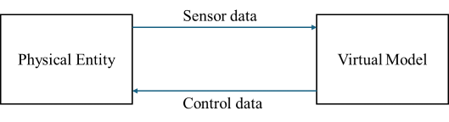}
	\caption{DT component.}
	\label{FIG2}
\end{figure}

In this work, we present a novel in-vehicle Digital Twin (DT)-based collision warning framework that detects and mitigates Sybil attacks by leveraging both road network geometry and C-V2X-based BSMs from surrounding vehicles. A DT model typically consists of three components: the physical entity, its virtual representation, and real-time data exchange between the two \cite{15} as illustrated in Fig. \ref{FIG2}. The in-vehicle DT-based framework presented in this work consists of a resource-efficient, analytical Sybil attack detection module and a DT-based collision warning module. The Sybil attack detection module leverages a Temporal Convolutional Network (TCN) encoder and a Hierarchical Navigable Small World (HNSW) classification algorithm. Together, the TCN encoder and the HNSW algorithm capture the spatio-temporal relationships among vehicle trajectories and classify them into Sybil (fake) and non-Sybil (real) categories in real-time. The DT-based collision warning module enables CVs to collect BSM data from nearby physical vehicles with C-V2X, replicate those vehicle trajectories in the in-vehicle virtual model using real-time data synchronization, and evaluate collision risks between vehicles. If the collision warning module detects an imminent collision, it communicates with the Sybil detection module to check if the lead vehicle is Sybil or non-Sybil. With both modules operating in parallel, the framework detects and mitigates Sybil attacks, thereby reducing Sybil-induced false collision warnings and unnecessary vehicle movement, leading to improved road safety. Using real-time data synchronization, our framework maintains a continuously updated virtual model of its surroundings within the DT, thereby providing robustness and continuity even when traditional sensors fail. By operating independently of RSU or backend infrastructure, our in-vehicle DT framework ensures continuous Sybil attack detection and mitigation capability even when RSU or infrastructure communication is ineffective, intermittent, or unavailable. While DT technology has seen growing adoption in the automotive sector for predictive maintenance, fleet management, and enhancing user experience \cite{15,16}, to the best of our knowledge, this is the first DT-based collision warning approach that can detect and mitigate Sybil attacks in CV environments.

The framework presented in this work is evaluated in the field using two distinct Sybil attack scenarios. In the first scenario, the attacker broadcasts BSMs for multiple fake vehicles that remain at a constant distance from the attacker vehicle. In the second scenario, the attacker broadcasts BSMs for fake vehicles whose distance offsets from the attacker vehicle vary dynamically. Two well-accepted surrogate measures of collision risk, time exposed time-to-collision (TET) and time integrated time-to-collision (TIT) have been used to evaluate the efficacy of the framework in improving road safety \cite{17,18,19}. TET quantifies the total time during which a vehicle experiences potentially unsafe car-following conditions by measuring the cumulative duration during which the time-to-collision (TTC) between a vehicle and its leader falls below a predefined safety threshold. In contrast, TIT aggregates the inverse of TTC over time, thereby assigning greater importance to situations with extremely small TTC values and reflecting the severity of potential collision events. TTC is the time to collision if the current speeds and trajectories of two vehicles are maintained. In addition, a real-world deployment feasibility study is also conducted to evaluate the framework’s latency and computational resource requirements. The evaluation of the framework yielded encouraging results, highlighting the potential of a DT-based approach as a robust, infrastructure-independent collision warning and security mechanism for CVs.

The key contributions of this study can be summarized as follows-

\begin{itemize} 
\item We introduce a DT-based, fully in-vehicle framework that jointly detects and mitigates Sybil attacks while enabling collision risk assessment without reliance on RSUs or backend infrastructure.
\item We develop a real-time, infrastructure-independent Sybil detection approach that leverages spatio-temporal vehicle trajectory modeling using a TCN encoder and HNSW classification. 
\item We validate the proposed framework through field experiments under static and dynamic Sybil attack scenarios, demonstrating reduced Sybil-induced collision risk as measured by TET and TIT.

\end{itemize}

The remainder of the paper is structured as follows. A review of related studies and the research gaps is presented in Section 2. Section 3 details the method followed for the framework. The dataset preparation process is discussed in Section 4. Section 5 presents the development of the in-house DT framework components. Section 6 entails the evaluation process, results and a brief discussion of the findings. Finally, Section 7 concludes the paper by summarizing the findings, limitations, and future directions.

\section{Related work}

In this section, we discuss prior research on Sybil attack detection and studies that employ DT–based systems for cybersecurity applications. The discussion highlights key methods and limitations in these two areas to contextualize and motivate the framework presented in this work.

\subsection{Application of DT in Cybersecurity and ITS Safety}

DTs are widely used for predictive incident research to ensure security across industries such as manufacturing, infrastructure management, power, and aerospace. In fact, DT has become one of the most exciting research concepts in autonomous and intelligent manufacturing owing to rapid advancements in mechanical, electrical and information technologies, as well as in virtual simulation and data acquisition technologies \cite{20}.

Although not as prominent as the industrial manufacturing process, DT has been used by researchers in the field of cybersecurity, particularly for attack detection and prevention. Masi et al. developed a framework for designing a security-oriented DT for cyber-physical critical systems, adopting a security-by-design architectural approach to simulate cyberattacks and develop corresponding countermeasures\cite{21}. The authors demonstrated their work leveraging the RAMI 4.0 industrial framework. Korman et al. evaluated the extent to which a DT-based countermeasure can improve the security of the Supervisory Control and Data Acquisition (SCADA) system of a power grid \cite{22}. A DT framework leveraging Long Short-Term Memory with Deep Reinforcement Learning (LSTM-DRL) model has been presented by Ali et al. to detect coordinated cyberattacks on electric vehicle grid cyber-physical systems \cite{23}. Eckhart et al. presented a framework called Cyber-Physical Systems (CPS) Twinning to generate DTs from the CPS’s physical entity specification \cite{24}. Using CPS Twinning, the authors presented a method to easily detect an intrusion with a comparison of specification-based DT signals and real device signals \cite{25}. A DT-based predictive method for anomaly/failure detection has been presented by Kummerow et al. This DT framework for the transmission system is called a dynamic digital model (DDM), which detects anomalies by analyzing current signals alongside historical data or trends and comparing them with measured signals \cite{26}.

DTs have recently gained prominence in the ITS and infrastructure domains for safety and security-related applications. References \cite{27,28,29} provide several examples of DT technology applied to railroads, highways, and bridges for design, monitoring, and maintenance, demonstrating significant impacts on safety, energy conservation, emission reduction, and efficient operation of real-world systems. Li et al. developed a Driver Risk-Aware Mobility Analytics DT that anticipates safety risks and demonstrated strong predictive accuracy \cite{30}. 

\subsection{Sybil Attack Detection}
Cyberattack on vehicles is a well-researched topic. Prior to 2015, most attacks on vehicles exploited vulnerabilities in the Controller Area Network (CAN) bus or Electronic Control Unit (ECU) through the physical On-Board Diagnostics (OBD) port, which required physical access to the vehicle \cite{31}. With the introduction of Bluetooth, cellular internet, and infotainment systems, studies have documented remote attacks on vehicles using different communication networks. Several research works have already demonstrated that communication networks, such as Dedicated Short Range Communication (DSRC) and cellular technologies, can be exploited for CVs. For instance, Petit et al. demonstrated risks to the connected and autonomous vehicles stemming from several attacks on Vehicle-to-Vehicle (V2V) communication, emphasizing the critical impact on system reliability and safety \cite{4}. One such cyberattack, Sybil attacks, has a critical impact on communication integrity, traffic safety, and network reliability. Consequently, Sybil attack detection in vehicular networks has attracted significant research interest. Research on Sybil attack detection in vehicular environments can be broadly classified into four categories: secure authentication-based methods, location-based methods, resource-testing-based methods, and data-driven behavior-measurement-based methods \cite{10}.

Yang et al. introduced a classification approach in which vehicle mobility behavior is analyzed using three classification algorithms at a centralized server \cite{11}. Data is collected via base stations that provide location certification, enabling the server to distinguish Sybil-generated fake vehicles by correlating certified locations with mobility patterns and by constructing co-occurrence graphs. However, centralized processing is essential for the model training and decision-making \cite{10}. Asad et al. developed a federated learning framework (FL-SATS) that leverages a three-tier architecture comprising vehicles, RSUs, and a software-defined network controller to detect Sybil attacks. Their method enables distributed model training at vehicles, with RSUs and central controllers aggregating model weights. This hierarchy balances detection accuracy, privacy, and low latency; however, it requires RSUs and centralized SDN controllers as key detection components \cite{32}. The collaborative learning method by Azam et al. uses ensemble ML classifiers on mobility data for Sybil detection. Although the focus is on algorithm robustness, data acquisition and model training are tied to centralized infrastructure or RSUs for gathering sufficient training data and distributing detection results, thereby emphasizing a semi-centralized architecture \cite{33}. The approach by Yao et al. leverages multi-channel monitoring using RSUs to detect Sybil nodes by identifying anomalies in channel access patterns. Since a Sybil attacker cannot transmit simultaneously on multiple channels with different identities, monitoring for such behavior anomalies and analyzing inconsistencies is used to detect attackers \cite{34}. Rakhi et al. presented a decentralized detection technique based on Longest Common Subsequence (LCSS) similarity and change-point detection at cluster heads (CHs) within vehicular clusters. Their cluster-based detection method compares the Received Signal Strength Indicator (RSSI) sequence to identify power-controlled Sybil attacks, thereby enhancing scalability and reducing reliance on infrastructure \cite{35}. Attar et al. analyzed Sybil attack detection using classification algorithms trained on mobility features extracted from vehicular data. Though focused on algorithmic performance, the data processing relies on centralized or semi-centralized gathering facilitated by RSUs or servers, maintaining typical infrastructure reliance \cite{36}. Baza et al. presented a Sybil attack detection mechanism that combines cryptographic proofs of location issued by multiple RSUs and proof-of-work puzzles to limit the creation of Sybil identities. Vehicles accumulate location proofs signed by RSUs, and detection is performed by a central event manager by analyzing trajectories. This approach offers strong privacy protections but relies on RSUs and centralized event management for authentication and Sybil separation \cite{37}. 

Although previous studies on DT enhance anomaly detection, operational efficiency, and risk analysis, none provide a real-time mechanism for detecting and mitigating attacks to safeguard CVs. Existing Sybil attack detection methods leverage RSUs for Sybil detection; however, the risk of signal attenuation or communication loss caused by real-world obstacles around the RSUs was not addressed in these studies. Furthermore, a centralized approach to Sybil attack detection can be a single point of failure for a region, affecting attack detection and mitigation for all CVs in the region. These gaps motivate our work: leveraging in-vehicle DT as an active attack mitigation mechanism for CVs under Sybil attack, while operating independently of RSUs, to ensure continuous and robust attack detection.

\section{Method}

This section introduces the Sybil attack model and the components of the in-vehicle DT-based collision warning framework. The subsequent subsections describe the attack models used for the field test, provide an overview of the DT-based collision warning architecture, and detail the mathematical foundations underlying its core components.

\subsection{Attack Model}

This work considers a Sybil attacker targeting a C‑V2X-supported CV system. The adversary is assumed to possess multiple valid vehicle identities (IDs) and to be capable of generating and transmitting forged BSMs under different vehicle IDs. 

There are four Sybil attack generation methods described in the literature, which are (i) generating completely random values for each BSM field, (ii) generating BSM packets with random values constrained by locally observed ranges from surrounding vehicles’ BSMs within a specific time frame, (iii) replaying captured BSM packets with modified timestamps and vehicle IDs, and (iv) constructing grid-pattern fake vehicles where the fake BSMs are generated based on the attacker vehicle’s position, speed, heading etc. \cite{38,39}. Random BSM generations using the first and the second methods produce uncorrelated trajectories that do not realistically place fake vehicles on a collision course with the victim, making them less effective for systematically attacking a C-V2X-based collision warning application. In the third method, an attacker rebroadcasts BSMs captured from other real vehicles after modifying the timestamp and vehicle IDs, so the attack depends on the natural motion of some real vehicles recorded earlier and only creates collision scenarios when their trajectories happen to intersect the victim vehicle’s path, limiting the attacker’s real-time control over the Sybil attack. In contrast, the fourth method, i.e., the grid-pattern-based method, enables the attacker to generate a set of fake vehicles around their own vehicle. The positions, speeds, and headings can be selected by the attacker so that multiple fake vehicles can be deliberately aligned on potential collision trajectories with the victim, leading the victim CV to believe it is about to collide with a vehicle. This method provides the attacker with real-time control over the spatial and temporal characteristics of the attack and is therefore well-suited for practical, effective attacks against a C-V2X-based collision warning application.

\begin{figure}
	\centering
	\includegraphics[width=\columnwidth]{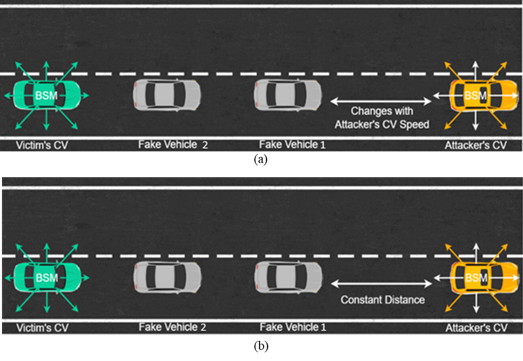}
    
	\caption{Sybil attack models of (a) Scenario 1, where the distance between the attacker CV and the fake vehicle is dynamic, and (b) Scenario 2, where the distance between the attacker CV and the fake vehicle is constant. }
	\label{FIG3}
\end{figure}

The following two scenarios of grid-pattern Sybil attacks were considered in this study for a two-prong assessment of the Sybil detection model developed in this study:

Scenario 1: In this attack scenario, the attacker vehicle broadcasts BSMs containing the location and speed of its own. Simultaneously, the attacker vehicle generates fake BSMs for multiple non-existent vehicles placed at a distance from itself. The distance between each fake vehicle and the real attacker vehicle varies with the attacker vehicle’s speed. When the attacker is driving slowly, the fake vehicles appear closer to the attacker vehicle, and as the attacker increases speed, the fake vehicles appear farther away. This creates moving, fake vehicles whose positions are dynamically controlled based on the attacker vehicle’s speed.

Scenario 2: Similar to the first attack scenario, the attacker vehicle broadcasts regular BSMs, as any CV would, and simultaneously generates fake BSMs for multiple nonexistent vehicles. But, unlike Scenario 1, the distance between the real attacker vehicle and each fake vehicle remains constant in Scenario 2. The fake vehicles always maintain a fixed distance from the real attacker vehicle, regardless of their operating speeds, creating similar vehicle motion patterns to that of the attacker vehicle while maintaining a consistent distance offset. This attack causes other CV systems to perceive several vehicles traveling in a platoon formation, even though only the attacker vehicle actually exists in reality.

These two scenarios are selected to capture both platoon-like, steady traffic patterns and more realistic, dynamically evolving traffic patterns, thereby enabling a two-prong assessment of the efficacy of the Sybil detection model developed in this study. Fig. \ref{FIG3} illustrates both of these grid-pattern Sybil attack scenarios. 

\subsection{In-Vehicle DT Framework Overviews}

In this study, the in-vehicle DT framework consists of three components: (a) the physical vehicles, equipped with sensing and communication devices and operating on a real-world road network; (b) a virtual layer, implemented on the victim vehicle’s onboard computer, that hosts the Sybil detection and collision warning modules, where the collision warning module maintains a digital representation of the road geometry and vehicles; and (c) a real-time communication channel that synchronizes information between the physical and virtual layers to enable continuous monitoring for Sybil attacks and potential collisions. 

\begin{figure}
	\centering
	\includegraphics[width=\columnwidth]{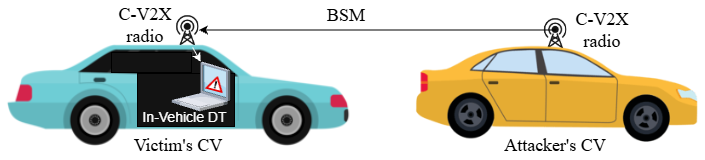}
	\caption{Experimental setup of in-vehicle DT model.}
	\label{FIG4}
\end{figure}

\begin{figure}
	\centering
	\includegraphics[width=\columnwidth]{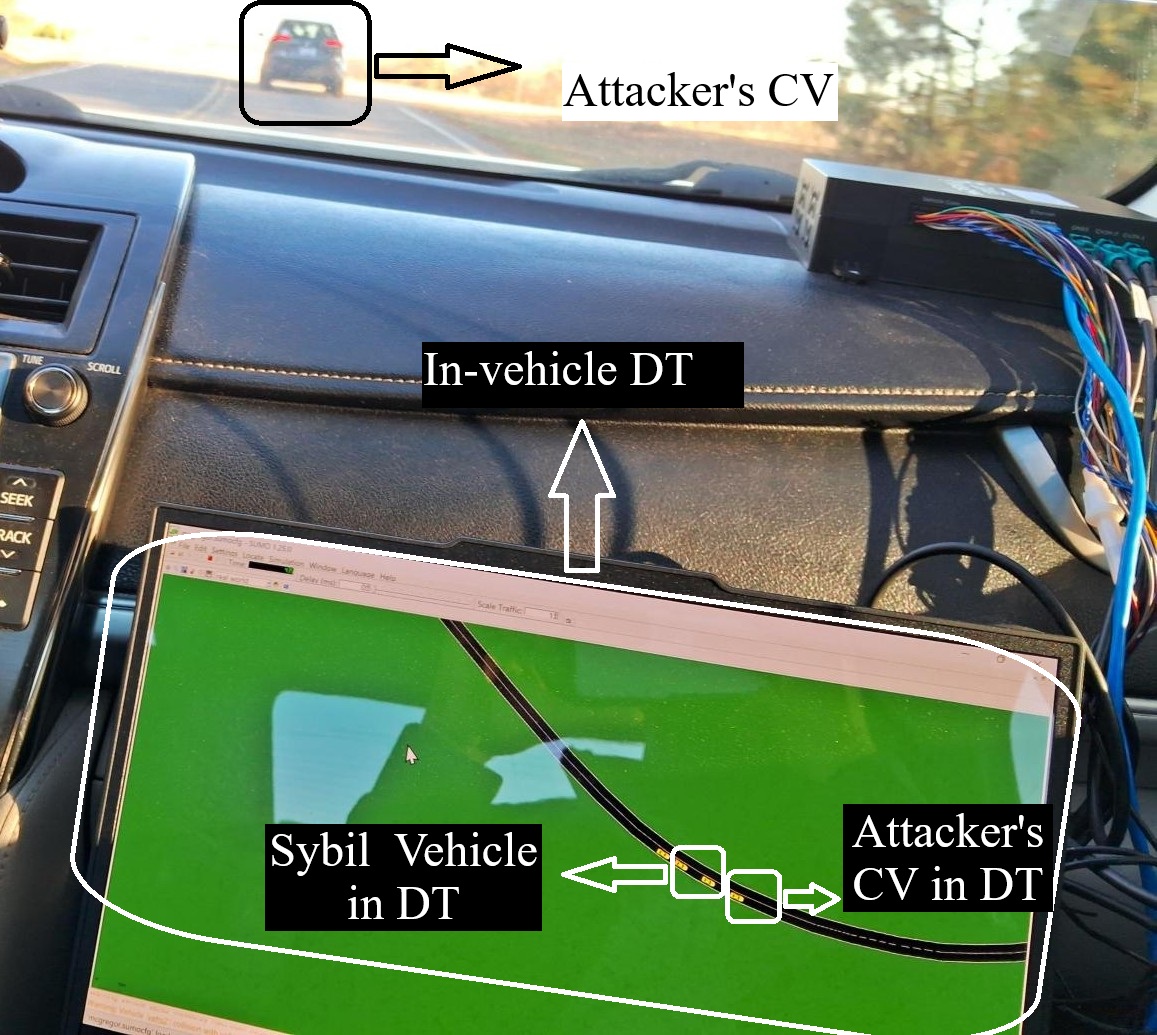}
	\caption{Experimental setup of the in-vehicle DT framework showing the attacker and fake vehicles inside DT model.}
	\label{FIG5}
\end{figure}

The physical vehicles are equipped with GPS receivers and C-V2X radios that support localization and direct V2V communication via the C-V2X PC5 interface, respectively. The CVs broadcast and receive BSMs at 100-millisecond intervals \cite{40}, in accordance with the Society of Automotive Engineers (SAE) J2735 standards \cite{7}. Fig. \ref{FIG4} demonstrates the attack scenarios considered in this study. The victim vehicle receives BSMs transmitted by the attacker and ingests them into the in-vehicle DT environment, where they are used to continuously update the virtual representation of the real-world traffic condition. Fig. \ref{FIG5} shows the in-vehicle DT framework during the field test, where the BSMs received from the attacker CV were used to replicate vehicle motion in the digital model of the road network.

Within the in-vehicle virtual layer, the collision warning module maintains an accurate digital model of the road network, including lane geometry and GPS-referenced coordinates, along with dynamically updated vehicle positions and trajectories. Incoming BSMs are ingested in real-time to reconstruct vehicle motion on this digitized network and to evaluate potential collision conditions. In parallel, the Sybil detection module processes the same BSM stream to identify inconsistencies or anomalous trajectory patterns indicative of Sybil behavior. 

\begin{figure}
	\centering
	\includegraphics[width=\columnwidth]{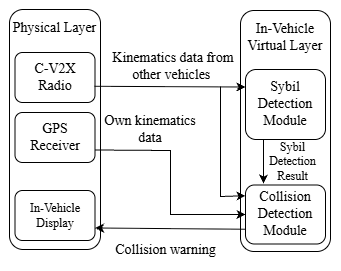}
	\caption{In-vehicle DT-based collision warning framework data flow.}
	\label{FIG6}
\end{figure}

The continuous bidirectional data exchange between the physical and virtual layers enables the in-vehicle DT to perform real-time collision risk assessment of the victim vehicle with real vehicles while identifying fake ones. When the virtual layer detects an imminent collision or flags a Sybil attack, it issues corresponding warnings to the physical vehicle, enabling on-time mitigation. The overall data flow of the in-vehicle DT framework is summarized in Fig. \ref{FIG6}.

\subsubsection{Sybil Attack Detection Module}

\begin{figure}
	\centering
	\includegraphics[width=\columnwidth]{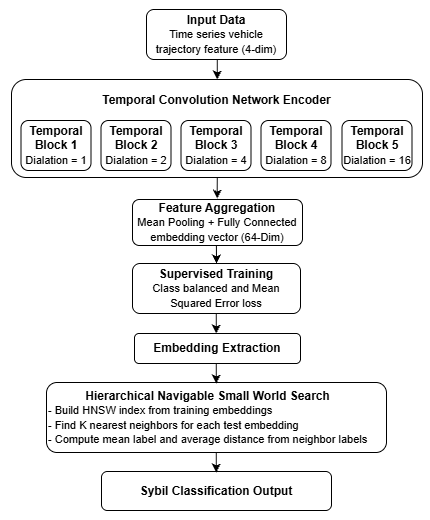}
	\caption{Sybil attack detection module workflow.}
	\label{FIG7}
\end{figure}

For the Sybil attack detection module, we present a hybrid ML model comprising an encoder and a similarity-based classification algorithm, which is presented in Fig. \ref{FIG7}. The Temporal Convolutional Network (TCN) encoder converts time-series vehicular trajectory data into a high-dimensional vector representation or embedding, efficiently learning and capturing the essential temporal patterns in each vehicle’s trajectory. Five temporal convolutional layers with increasing dilation factors extract multiscale temporal features from the 4-dimensional inputs: latitude, longitude, speed, and heading. These four input features are selected because they are fundamental components of standard BSMs. The resulting sequence representations are aggregated via mean pooling and a fully connected layer to form fixed-size embedding vectors, which are then used in supervised training with a class-balanced strategy and mean squared error loss to learn discriminative embeddings for Sybil detection. After training, embeddings for test trajectories are extracted and indexed in an HNSW graph built from the training embeddings to infer Sybil labels based on neighbor classes and their average distances. This hybrid design was developed to leverage the strengths of both TCN encoders and HNSW. TCN effectively models the temporal dependencies and dynamics of trajectory sequences \cite{41}, while HNSW provides a fast, scalable mechanism for identifying suspiciously similar trajectories across the network \cite{42}.

Let $\mathbf{X}\in\mathbb{R}^{T\times d}$  denote the input sequence for a single vehicle, where $T$ is the sequence length, which is the number of BSM data points collected within a time window and $d=4$ is the number of features (i.e., latitude, longitude, speed, heading). Latitude, longitude and heading are normalized using mean normalization:
\begin{equation}
\begin{aligned}
X_{:,j|norm}=\frac{X_{:,j}}{\mu_j},\ j=1,\ldots,d-1
\end{aligned}
\end{equation}

where $\mu_j$ is the mean of the feature $j$ over the training set.

Speed was normalized by the posted speed limit to represent vehicle movement relative to the posted speed limit. This approach enables a consistent comparison of driving behavior on roads with different speed limits.
The TCN consists of $L=5$ temporal blocks, each employing a 1D dilated causal convolution. For block $l$, the output at time $t$ is given by:
\begin{equation}
\begin{aligned}
y_t^{\left(l\right)}=\sum_{i=0}^{k-1}w_i^{\left(l\right)}\cdot x_{t-d_l\cdot i}
\end{aligned}
\end{equation}
where $k=5$ is the kernel size, $d_l=2^{l-1}$  is the dilation factor for block $l$, and $w_i^{\left(l\right)}$ denote the learned convolutional weights. Zero-padding is applied to ensure that the output sequence length matches the input. The choice of $l$ and $k$ is determined by the sequence length that will be used as input. Each temporal block includes a residual connection:
\begin{equation}
\begin{aligned}
y^{\left(l\right)}=\mathrm{ReLU}\left(\mathrm{Dropout}\left(\mathrm{Chomp}\left(\mathrm{Conv1D}\left(x^{\left(l\right)}\right)\right)\right)+x^{\left(l\right)}\right)
\end{aligned}
\end{equation}
where ${x}^{\left(l\right)}$  is the input to the block $l$, and the residual is downsampled via a $1\times1$ convolution if the input and output dimensions differ. After the final block, global average pooling is applied over the temporal dimension:
\begin{equation}
\begin{aligned}
z=\frac{1}{T}\sum_{t=1}^{T}y_t^{\left(L\right)}
\end{aligned}
\end{equation}

The pooled vector ${z}\in\mathbb{R}^h$ (with $h=64$)  is then projected to the embedding space via a fully connected layer:
\begin{equation}
\begin{aligned}
e=W_{\mathrm{fc}}z+b_{\mathrm{fc}}
\end{aligned}
\end{equation}
where ${W}_{\mathrm{fc}}\in\mathbb{R}^{64\times h}$ and ${b}_{\mathrm{fc}}\in\mathbb{R}^{64}$.

The encoder is trained using a mean squared error (MSE) loss:
\begin{equation}
\begin{aligned}
\mathcal{L}_{\mathrm{MSE}}=\frac{1}{N}\sum_{i=1}^{N}|e_i-\widehat{e_i}|^2
\end{aligned}
\end{equation}
where ${\hat{\mathbf{e}}}_i$  is a 64‑dimensional target embedding generated from the binary label of the sequence $i$, by repeating the label value across all embedding dimensions.

The embeddings generated by the TCN are indexed using the HNSW graph, an efficient algorithm for approximate nearest neighbor search. The HNSW builds a layered proximity graph over the embedding space, enabling rapid and scalable similarity comparisons between generated embeddings. During inference, a new embedding is queried against the HNSW index containing previous trajectory embeddings. 

Let $\mathcal{E}_{\mathrm{train}}=\left\{\mathbf{e}_1,\ldots,\mathbf{e}_N\right\}$ denote the set of embeddings from the training set. The HNSW index organizes these vectors in a multi-layered graph structure to enable efficient approximate nearest neighbor search in the Euclidean space ($L_2$ metric). The index is constructed with the following parameters: $M$ denotes the number of bidirectional links per node (set to $16$), ${\mathrm{ef}}_{\mathrm{construction}}$  denotes the construction-time search depth (set to $16$), and $Ef$ denotes the query-time search depth (set to $50$).

Given a test embedding $\mathbf{e}_{\mathrm{test}}$, the HNSW index returns the indices and distances of the $k=3$ nearest neighbors:

\begin{equation}
\begin{aligned}
\{\left(i_j,d_j\right)\}_{j=1}^k=\mathrm{HNSW}\left(e_{\mathrm{test}},k\right)
\end{aligned}
\end{equation}
where $d_j=\parallel{e}_{\mathrm{test}}-{e}_{i_j}\parallel_2$.

HNSW finds the nearest neighbors in the embedding space and classifies the new embeddings in binary classes, i.e., Sybil or non-Sybil, based on their neighbors' embeddings.

Let ${y_{i_j}}_{j=1}^k$ be the labels of the $k$ nearest neighbors and let $\bar{d} = \frac{1}{k} \sum_{j=1}^{k} d_j$ be the average distance. The predicted label ${\hat{y}}_{\mathrm{test}}$  is assigned as:

\begin{equation}
\begin{aligned}
\hat{y}_{\mathrm{test}} =
\begin{cases}
0, & \text{if } \frac{1}{k}\sum_{j=1}^{k} y_{ij} < 0.5 \text{ and } \bar{d} > d_{\mathrm{threshold}}, \\
1, & \text{otherwise}.
\end{cases}
\end{aligned}
\end{equation}
where $0$ denotes a real or non-Sybil vehicle, $1$ denotes a Sybil attack-generated fake vehicle and $d_{threshold}$ is the $10^{th}$ percentile of all the average distances.

\subsubsection{DT-based Collision Warning Module}
The DT-based collision warning module presented in this study was designed as a resource efficient safety module that processes BSM data from the nearby CVs inside a DT model. The DT model consists of a digital replica of the real-world road network and the vehicles’ digital trajectories reconstructed from the BSMs.  From the digital trajectories, the module can identify other vehicles within the victim (inside which the DT is operational) vehicle’s proximity and generate an alert when the TTC falls below a predefined threshold. By correlating the output from the Sybil detection module, the collision warning module generates a collision alert if a collision risk is detected, and the Sybil detection module classifies the associated vehicles as non-Sybil. The collision warning module calculates the great-circle distance between two vehicles using the Haversine formula \cite{43}, as given in (9), which provides accurate distance estimation on the Earth’s surface based on geodetic coordinates. The distance between two vehicles d by the Haversine formula is- 
\begin{equation}
\begin{aligned}
d = R \times c
\end{aligned}
\end{equation}
where $c$ is the central angle given by:
\begin{equation}
\begin{aligned}
c\ =\ 2\ \times\ atan2\left(\sqrt a,\ \sqrt{\left(1-a\right)}\right)
\end{aligned}
\end{equation}
The GPS coordinates (latitude $\varphi$, longitude $\lambda$) between two vehicles is used to calculate the Haversine component $a$ given by:
\begin{equation}
\begin{aligned}
a\ =\ sin^2\left(\frac{\mathrm{\Delta\varphi}}{2}\right)+\cos{\left(\varphi_1\right)}\times\cos{\left(\varphi_2\right)}\times\ sin^2\left(\frac{\mathrm{\Delta\lambda}}{2}\right)
\end{aligned}
\end{equation}
where $\mathrm{\Delta\varphi}\ =\ \varphi_2-\ \varphi_1$,  $\mathrm{\Delta\lambda}\ =\lambda_2-\ \lambda_1$ and, $R$ represents the Earth’s mean radius ($R = 6,371$ kilometers). $(\varphi_1, \lambda_1)$ and $(\varphi_2, \lambda_2)$  represent the GPS coordinates of two real vehicles.
The TTC is a well-established parameter in collision warning applications, which is given by: 
\begin{equation}
\begin{aligned}
TTC=\ \frac{d}{V_{rel}}
\end{aligned}
\end{equation}
where $V_{rel}$ is the relative speed between the vehicles. When the TTC value falls below a threshold, the collision warning module issues a warning to the driver.

\section{Dataset}

This section describes the dataset used for the Sybil detection model development and evaluation, including data collection, processing and structuring for analysis. The first subsection outlines the experimental setup and data collection procedure, and the following subsection details the data preparation process.

\subsection{Experimental Setup and Data Collection}

Two CVs were driven along McGregor Road in Clemson, SC, during a field test to collect Sybil attack data. The attacker vehicle was positioned in front, broadcasting its own legitimate BSMs and fake BSMs for multiple fake vehicles. The victim vehicle was following behind the attacker. The attacker adjusted the positions of the fake vehicles so they appeared between the attacker and the victim vehicles. At each time step (at a 10 Hz rate), the attacker sent out multiple sets of BSMs using a C-V2X radio: one with the real attacker vehicle’s ID and others with fake vehicle IDs corresponding to the fake vehicles behind it. The victim vehicle received these BSMs, each appearing to be from a distinct vehicle. As a result, the C-V2X-based collision warning system in the victim vehicle generated alerts when it detected close proximity to the fake vehicles, even though the victim vehicle was actually at a safe distance from the real attacker vehicle. To generate sufficient data for model training and to introduce variability in the generated Sybil attack patterns, three cases were considered based on the number of fake vehicles generated by the attacker: Case 1, where one fake vehicle is generated; Case 2, where two fake vehicles are generated; and Case 3, where three fake vehicles are generated. The field experiment was conducted 13 times in total, comprising 4 runs for Case 1, 4 runs for Case 2, and 5 runs for Case 3. For each case, 2 runs were conducted with Sybil attack scenario 1 and the rest with Sybil attack scenario 2, as described in Section 3.1. During the experiment, 4,990 BSM data points for fake vehicles and 4,441 BSM data points for real vehicles were recorded. These BSMs were used to train and evaluate the Sybil attack detection model. Fig. \ref{FIG5} shows the DT simulation during the attack scenario, where the in-vehicle DT model shows four vehicles on the road, whereas in reality, there are only two.

The Sybil attack experiments required instrumented CVs, multiple test drivers, controlled safety procedures on public roads, and were time-consuming. Because of these reasons, it was only feasible to collect a limited amount of data for training. Hence, real-world driving data was augmented with data generated by the Simulation of Urban Mobility (SUMO) microscopic traffic simulation platform. Simulation-based data augmentation enabled training the sybil attack detection module across a range of traffic flows (i.e., 30\%, 60\%, and 90\% of the road capacity) while preserving vehicle operating characteristics similar to those of real-world vehicles and road network geometry of the field test site. 

The virtual road network for simulation was derived from OpenStreetMap (OSM) data for McGregor Road, located within Clemson University, Clemson, South Carolina (Universal Transverse Mercator Zone 17N). A route was created along McGregor Road using the OSM map data for simulation purposes. SUMO’s \textit{netconvert} utility was used to convert the OSM data into a SUMO-compatible network file. To simplify the topology and reduce simulation complexity, non-essential features, such as buildings, sidewalks, minor service roads, and pedestrian crossings, were excluded during network development. The network topology was verified using SUMO-GUI's visualization tools, ensuring accurate representation of (i) intersection geometries, (ii) lane configurations, and (iii) traffic signals. 

Each simulation run was for 20 minutes. The simulations represented vehicle flows of 510, 1020, and 1530 passenger cars, corresponding to 30\%, 60\%, and 90\% of a two-lane highway capacity under base conditions, respectively, as specified in the Highway Capacity Manual (HCM) 2010 \cite{44}. These capacities were selected to evaluate the Sybil detection model's performance under varying traffic densities, including sparse, moderate, and congested conditions. After the simulation, 15,826, 38,015, and 58,342 BSM data points were recorded for traffic flows at 30\%, 60\%, and 90\% of road capacity, respectively. 

\subsection{Data Preparation}

\begin{figure}
	\centering
	\includegraphics[width=\columnwidth]{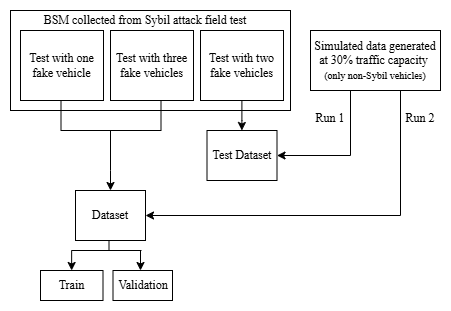}
	\caption{Dataset creation process from field test BSM and simulated data generated for traffic flow at 30\% of road capacity.}
	\label{FIG8}
\end{figure}

The datasets for training and validating the Sybil detection model were created by appending the real and fake vehicle BSMs collected during the Sybil attack field test to the simulation-generated data under varying traffic densities. Fig. \ref{FIG8} illustrates the dataset preparation process for traffic flow at 30\% capacity as an example; the same process was followed for the datasets for traffic flows at 60\% and 90\% capacity. First, BSMs for the real and fake vehicles collected from Sybil attack field experiments (with a single fake vehicle and three fake vehicles) were combined and appended to the simulated data generated for traffic flow at 30\% capacity to create the dataset. This dataset is further split into the training and validation sets. This process is repeated with simulated data for traffic flows at 60\% and 90\% capacity to create additional training and validation datasets. These training and validation datasets were used to train the Sybil detection model and to validate its performance for our framework development. A test dataset is created by appending the attacker, victim, and fake vehicle BSM collected during the field test with two fake vehicles to simulated data generated for traffic flow at 30\% capacity, as illustrated in Fig. \ref{FIG8}. The simulated data used in the test dataset is different from that of the train and validation datasets, generated using a different simulation run. The test dataset is used to evaluate the effectiveness of the in-vehicle DT framework.

In the datasets, binary classification is used to classify vehicles as Sybil or Non-Sybil. The attacker and victim vehicles driven on the road during the field test were labeled as non-Sybil vehicles, while the fake vehicles were labeled as Sybil. All the vehicles from the simulation-generated data were labeled as non-Sybil.  To determine the necessary temporal context for effective Sybil attack detection, the trajectory sequence length for each vehicle was considered, which is the number of consecutive BSM data points from a vehicle. A sequence length of $T$ represents $T/10$ seconds of BSM data as the vehicles broadcast 10 BSMs per second, providing $T/10$ seconds of temporal context. We experiment with sequence lengths of 10, 20, 30, 40, 50, 60, 70, 80, 90, and 100 samples to assess the Sybil detection model’s performance across varying temporal contexts. The number of embeddings generated for  each Sybil class for various sequence lengths is given by:
\begin{equation}
\begin{aligned}
S\ =\sum_{i=1}^{n}\left(N_i-\ T\ +\ 1\right)
\end{aligned}
\end{equation}
where $N_i$ is the number of BSM data points for a vehicle $i$ in the class, $n$ is the number of vehicles in the class and $T$ represents the sequence length.

\section{Development of the In-vehicle Digital Twin}

In this section, we present the details of our in-vehicle DT framework development. First, we describe the Sybil detection model's parameter selection, training, and testing process, and present a brief discussion of the validation results, based on which we selected the optimal parameters. In the second subsection, we present our DT-based collision warning module and the parameters used for the module. 

\subsection{Sybil Attack Detection Module}

A series of experiments was conducted to determine the best-performing Sybil attack detection model across varying traffic conditions, using the training and validation datasets described in Section 4.2. The training datasets were used to train the Sybil attack detection model for 10, 20, 30, 40, 50, and 60 epochs. The model used a learning rate of $0.001$, a batch size of $32$ and the ADAM optimizer during training. Since a large number of normal driving samples were generated via simulation to augment the limited field data, the resulting dataset contains substantially more benign than Sybil attack instances. Consequently, the binary classes (i.e., Sybil and Non-sybil) are imbalanced, and a standard class-balancing technique, WeightedRandomSampler, is employed to ensure that each mini-batch drawn during training is class-balanced. The technique assigns a weight to each sample based on its class, ensuring that samples from underrepresented classes are sampled more frequently. However, the loss function itself is not class-weighted, and balancing is achieved only through the sampling strategy. Introducing class weights in the loss function would double-compensate for the imbalance, potentially biasing the model toward the minority class and leading to unstable training or degraded generalization performance.

After training, each model was validated on its corresponding validation dataset, and performance was measured using standard binary classification metrics, including accuracy, precision, recall, F1-score, and the area under the receiver operating characteristic curve (ROC-AUC). The best model was selected for testing based on the highest F1 score. 

\begin{figure*}[b]
	\centering
	\includegraphics[width=\textwidth]{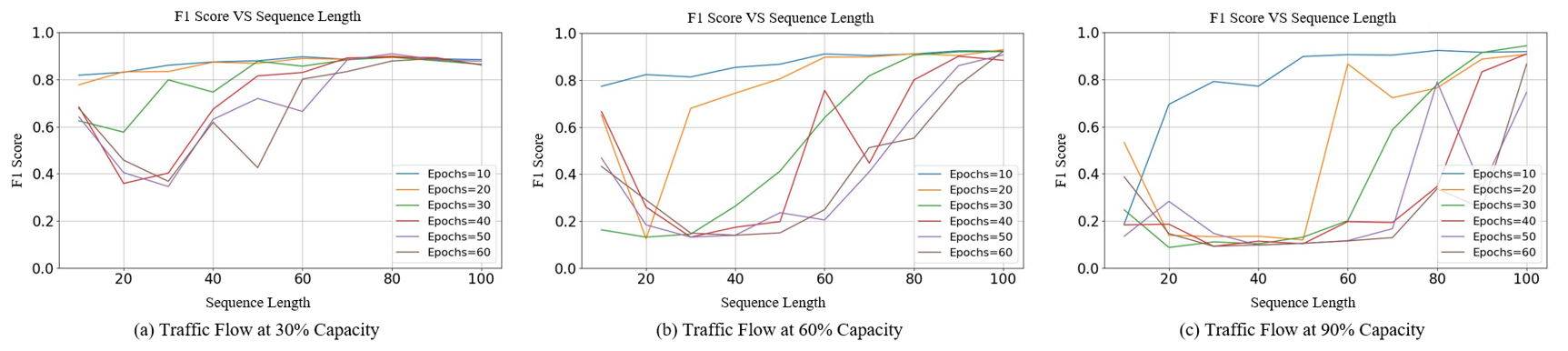}
	\caption{F1 Score performance comparison of the Sybil detection model for traffic flows at (a) 30\% capacity, (b) 60\% capacity, (c) 90\% capacity over varying sequence lengths and epochs.}
	\label{FIG9}
\end{figure*}

Fig. \ref{FIG9} illustrates the F1 score performance of the Sybil attack detection model with varying traffic flows, sequence length, and epochs. The results demonstrate that the F1 score is sensitive to the sequence length across all traffic densities. The model requires longer sequence lengths to improve the F1 score, which converges to a range of $0.90-0.94$ across various traffic densities. Shorter sequence lengths result in lower, more variable F1 scores across training epochs. This can be attributed to insufficient temporal context data to effectively distinguish between Sybil and non-Sybil vehicles. In sparse traffic, many benign vehicles follow similar near–speed–limit speed profiles over short horizons, resulting in very similar vehicle motion patterns. In moderate to dense traffic conditions, vehicles begin to follow each other closely and at short sequence lengths, and many vehicles’ motion patterns look similar. This leads some vehicles to fall into Sybil-dominated neighborhoods in the HNSW classifier and be flagged as false positives, resulting in lower F1 scores across all traffic densities. Higher training epochs with insufficient temporal context led to more false positives, further lowering the F1 score. As the sequence length increases beyond 70, most models’ F1 scores gradually increase towards convergence. Therefore, longer sequence lengths provide the model with sufficient context to classify borderline real vehicles, thereby reducing false positives and increasing the F1 score. The best F1 scores across all traffic conditions were observed when the sequence length was between 80 and 100. Further increasing the sequence length beyond 100 did not improve the F1 score during the experiments, indicating that the required temporal context for optimum model performance has already been captured, and data from earlier timestamps may not provide additional learning context.
The figure indicates that with longer sequence length, the model performs well across traffic densities with only minor improvements at higher densities. This can be attributed to the availability of more vehicle data in the training dataset at higher traffic densities. Furthermore, across all traffic flow conditions and sequence lengths, the Sybil detection models achieved their best performance within the first 50 training epochs. Beyond this point, additional training did not yield any improvements in the F1 score. This pattern suggests that the models converge relatively quickly and that prolonged training might lead to overfitting to trajectory-specific noise rather than learning new discriminative temporal patterns. Consequently, limiting training to at most 50 epochs provides an effective trade-off between detection performance and computational cost. Table \ref{table1} summarizes the performance metrics of the best-performing models across different traffic flows. Comparing the results, the model performed best when trained on the dataset prepared for traffic flow at 90\% capacity, with a sequence length of 100 and for 30 epochs.

\begin{table*}[t]
\centering
\small

\caption{Metrics for the best-performing model for traffic flows at 30\%, 60\% and 90\% of road capacity.}
\label{table1}
\renewcommand{\arraystretch}{1.15}
\setlength{\tabcolsep}{8 pt}

\begin{tabular}{>{\centering\arraybackslash}p{0.2\linewidth}ccccccc}
\toprule
\textbf{Traffic flow (\% of capacity)}& \textbf{Sequence Length}& \textbf{Epoch}& \textbf{Accuracy}& \textbf{Precision} & \textbf{Recall}& \textbf{F1}&\textbf{ROC-AUC}\\

30\%& 80& 50& 0.946& 0.843& 0.989& 0.910&0.959\\

60\%& 100& 20& 0.971& 0.872& 0.996& 0.930&0.980\\

90\%& 100& 30& 0.984& 0.893& 1.000& 0.944&0.991\\
\end{tabular}
\end{table*}

\subsection{DT-based Collision Warning Module}

The collision warning module used in this study is a DT-based module designed using SUMO. A digital replica of the McGregor road in Clemson is created as described in Section 4.1. A Python-based control script injects BSM data into the SUMO simulation in real-time via the Traffic Control Interface (TraCI) API, enabling real-time interaction with the simulation to ingest or export data. The DT model continuously computes the positions and velocities of other vehicles relative to the victim vehicle at a 100-millisecond interval to evaluate the collision risk based on a predefined critical TTC. This interval is used because the C-V2X radio broadcasts BSMs at a 100-millisecond interval as required by SAE J2735 [7]. When the TTC falls below the critical threshold for a vehicle in front of the victim vehicle (in which DT is running), the collision warning application checks the Sybil class associated with that leading vehicle, as identified by the Sybil detection module. If the leading vehicle is classified as non-Sybil, the collision warning module generates an alert; otherwise, the alert is suppressed. For this work, a critical TTC value of 2 seconds has been selected based on previous studies \cite{45}.

\section{Evaluation}

To assess the effectiveness of our framework, we conduct an evaluation of collision risks across three conditions: (i) baseline traffic with no attack, (ii) traffic under Sybil attack without our mitigation framework, and (iii) traffic under Sybil attack with our mitigation framework enabled. A traffic flow with preconfigured vehicle mobility settings was simulated in SUMO to represent the traffic condition with no attack. While a SUMO simulation was running, the recorded BSMs from the test dataset, described in Section 4.2, were injected to emulate the presence of Sybil vehicles, representing traffic under attack in the absence of our mitigation framework. This setup allowed the simulated vehicles to interact with the injected Sybil vehicles. Finally, the same simulation with Sybil vehicles was run again while our mitigation framework was active. Traffic flows corresponding to 30\%, 60\%, and 90\% of the roadway capacity were considered to assess the effect of the Sybil attack and the effectiveness of our mitigation framework across traffic densities.

\subsection{Evaluation Metrics}

Two surrogate measures for evaluating collision risk, i.e., TET and TIT, have been considered during the experiments. TET is defined as the total simulation time during which TTC is below a chosen threshold. For a vehicle $i$, TET can be presented as
\begin{equation}
\begin{aligned}
\mathrm{TET}_i
=
\sum_{k}^{}
\delta_{i,k}\,\Delta t,
\qquad
\delta_{i,k}=
\begin{cases}
1, & 0<TTC_{i,k}\le TTC_{\mathrm{threshold}},\\
0, & \text{otherwise}.
\end{cases}
\end{aligned}
\end{equation}
where $k$ is the timestep index, $\delta_{i,k}$ is an indicator variable that equals $1$ when the TTC of the vehicle $i$ at time step $k$ is below the specified threshold and 0 otherwise,  $\Delta t$ is the simulation step and ${TTC}_{Threshold}=2$ seconds.

TIT measures the cumulative severity of potential collision events by integrating the inverse of TTC over time and can be represented by the following equation. For a vehicle $i$,
\begin{equation}
\begin{aligned}
{TIT}_i\ =\sum_{k:{TTC}_{i,k}<\infty}{\frac{1}{{TTC}_{i,k}}\Delta t}
\end{aligned}
\end{equation}
where $k$ is the timestep index and $\Delta t$ is the simulation step.

To prevent numerical instability caused by extremely small TTC values, TTC was clipped at a minimum value of 0.1 s before computing TIT. This ensures that individual time steps do not produce disproportionately large contributions while still capturing the severity of near-collision events.

In addition, the framework is evaluated in terms of end-to-end latency and computational resource consumption to assess its suitability for real-time, in-vehicle deployment.

\subsection{Evaluation Result and Discussions}

In this section, we present and discuss the effectiveness of our in-vehicle DT-based framework using the surrogate safety metrics outlined in Section 6.1. During the experiments, the surrogate safety metrics TET and TIT are computed for all vehicle pairs to quantify collision risks between them when TTC is below 2 seconds. We also present latency analysis and computational resource consumption analysis of the framework to evaluate its feasibility for real-world deployments.  

\subsubsection{TET Analysis}

\begin{figure}
	\centering
	\includegraphics[width=\columnwidth]{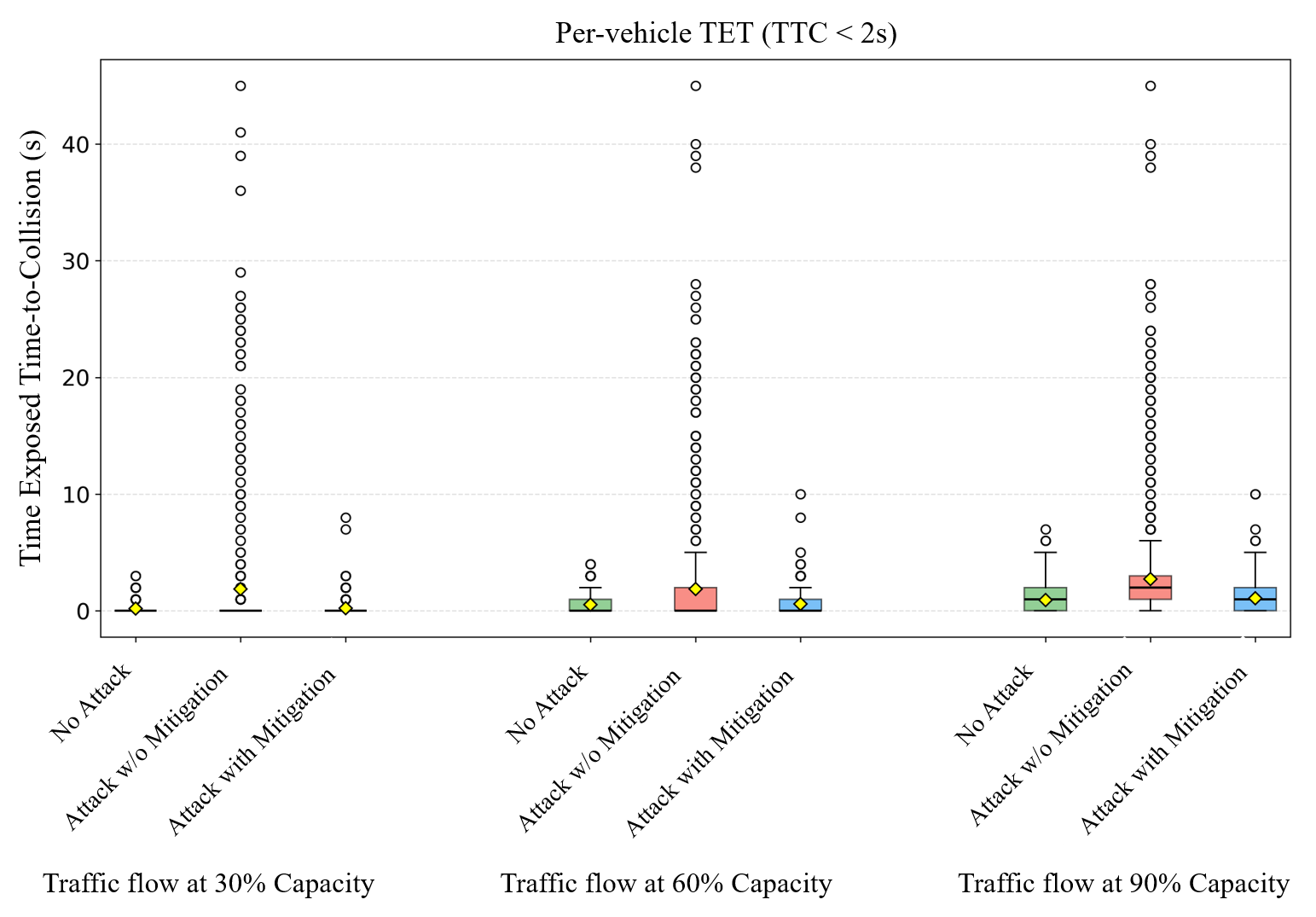}
	\caption{Box plot diagram of TET during no attack, attack without mitigation and attack with mitigation conditions at different traffic flows.}
	\label{FIG10}
\end{figure}

TET measures the total duration a vehicle spends with TTC below the chosen safety threshold of 2 seconds. The boxplot of per-vehicle TET at different traffic flows in Fig. \ref{FIG10} illustrates that Sybil attacks substantially increase the vehicles’ exposure to risky conditions, and the DT-based detection framework reduces exposure to near-baseline conditions.

Under attack conditions without mitigation, there is a noticeable increase in the mean TET from 0.180, 0.526, and 0.941 seconds in the no attack condition to $1.860$, $1.881$, and $2.715$ seconds for traffic flows at 30\%, 60\%, and 90\% capacity, respectively. This can be attributed to the Sybil attack-induced unsafe condition, where more vehicles are exposed to dangerous conditions for a longer period. A noticeable increase in extreme outliers is also evident in the figure, further confirming that some vehicles experience long exposure to near-collision events during the Sybil attack. 

Applying our DT-based mitigation framework consistently reduces the mean TET across all traffic flows and conditions. Fig. \ref{FIG10} demonstrates that the attack with mitigation framework conditions has Interquartile Ranges (IQRs) and means much closer to the baseline condition, with fewer outliers, indicating that detection prevents many of the long-duration exposures caused by sybil injections. The mean TET in these conditions is $0.228$, $0.583$, and $1.071$ seconds for traffic flows at 30\%, 60\%, and 90\% of road capacity, respectively, representing reductions of 88\%, 69\%, and 60\% compared to the attack without mitigation condition. The small number of residual outliers remains under attack with the mitigation condition. Although the Sybil vehicles were removed from the in-vehicle DT, the real attacker vehicle likely caused traffic congestion due to its lower speed during the attack.
A paired t-test with a significance level of $0.005$ was conducted to evaluate the effect of Sybil attacks and the effectiveness of our mitigation mechanism on TET across 30\%, 60\%, and 90\% traffic densities. The analysis compares TET values of the same set of vehicles under different conditions, which are (i) No Attack versus Attack without mitigation, and (ii) Attack without mitigation versus Attack with mitigation. A paired t-test was used to account for the dependency between observations from identical vehicles across scenarios. Two directional hypotheses were formulated. For the first comparison, the null hypothesis, $H_0$, assumes that the mean TET during attack does not increase, i.e., the attack does not increase exposure to unsafe conditions, while the alternative hypothesis, $H_1$, tests whether the attack leads to higher mean TET. The hypotheses can be represented as

$\begin{aligned}
H_0:\mu_{\mathrm{No\ Attack}}\geq\mu_{\mathrm{Attack\ without\ mitigation}}
\end{aligned}$

$\begin{aligned}
H_1:\mu_{\mathrm{No\ Attack}}<\mu_{\mathrm{Attack\ without\ mitigation}}
\end{aligned}$

For the second comparison, the null hypothesis $H_0$ assumes that mitigation does not reduce the mean TET, while the alternative hypothesis $H_1$ evaluates whether the mitigation mechanism lowers the mean TET. For the second comparison, the hypotheses can be represented as

$\begin{aligned}
H_0:\mu_{\mathrm{Attack\ without\ mitigation}}\le\mu_{\mathrm{Attack\ with\ mitigation}}
\end{aligned}$

$\begin{aligned}
H_1:\mu_{\mathrm{Attack\ without\ mitigation}}>\mu_{\mathrm{Attack\ with\ mitigation}}
\end{aligned}$

For both comparisons, the p-values are extremely small across all traffic conditions, ranging from ${10}^{-11}$to ${10}^{-61}$, leading to rejection of both null hypotheses. These findings indicate that the Sybil attacks significantly increased the duration of unsafe conditions, while the mitigation mechanism significantly reduced this exposure. The consistency of these results across varying traffic densities further supports the robustness of the observed effects.

\subsubsection{TIT Analysis}

\begin{figure}
	\centering
	\includegraphics[width=\columnwidth]{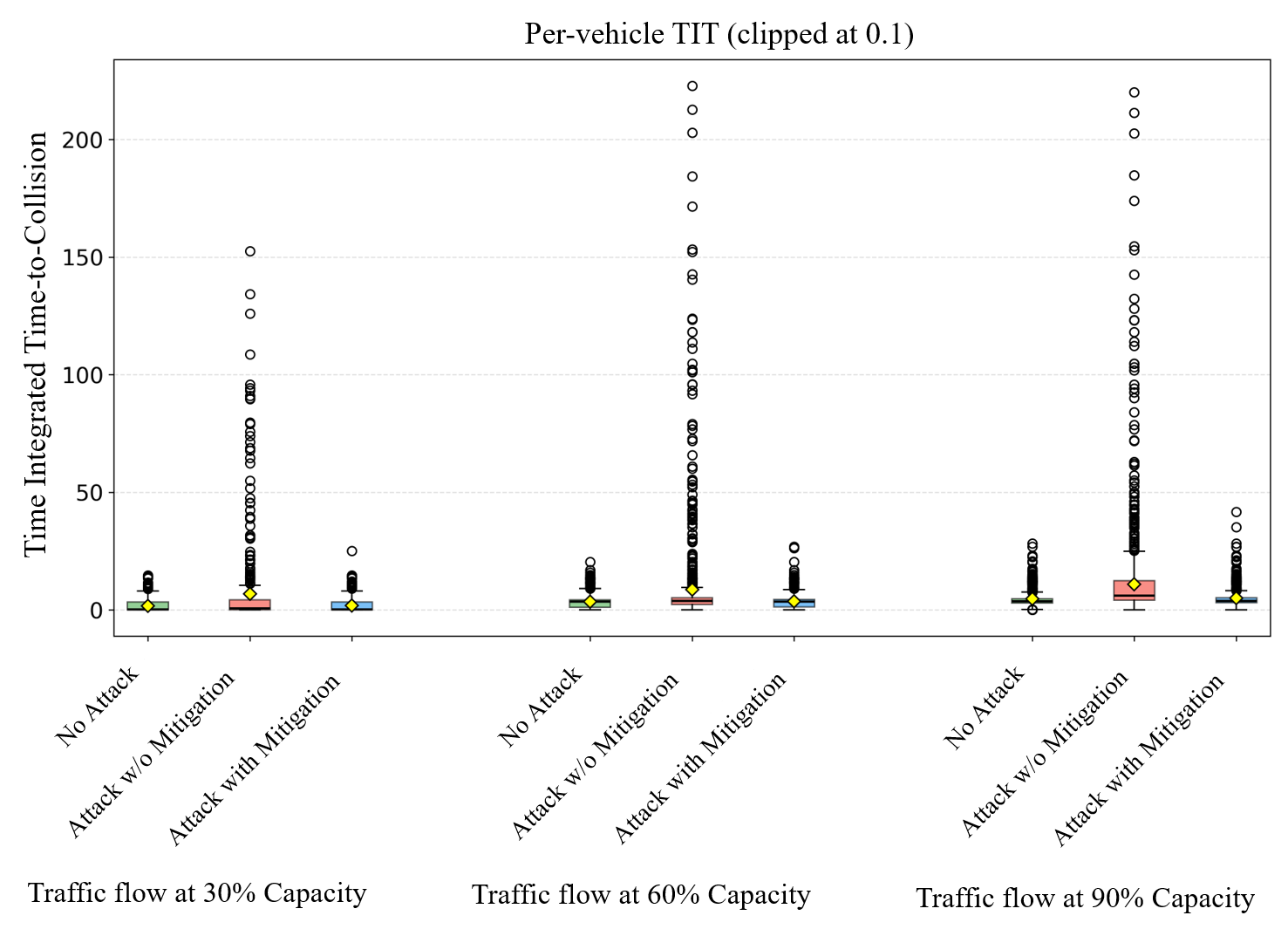}
	\caption{Box plot diagram of TIT during no attack, attack without mitigation and attack with mitigation conditions at different traffic flows.}
	\label{FIG11}
\end{figure}

To further evaluate the severity of unsafe traffic interactions, the TIT metric was analyzed. Unlike TET, which measures the duration of unsafe conditions experienced by the vehicles, TIT captures the severity of those conditions by assigning greater weight to situations with extremely small TTC values. A box plot of the TIT distribution for different attack conditions and various traffic flows is presented in Fig. \ref{FIG11}. The results indicate that Sybil attacks significantly increase TIT values across all traffic densities, reflecting a higher exposure to severe near-collision events.

At all traffic flows, the increase in mean TIT is particularly noticeable, rising from $1.792$, $3.551$, and $4.734$ in the no attack condition to $6.786$, $8.686$, and $10.970$ during the attack without mitigation for traffic flows at 30\%, 60\%, and 90\% capacity, respectively. In high-density traffic conditions, sudden braking due to a fake vehicle on the road can lead to critical TTC, as there may not be sufficient space for the follower vehicles, resulting in a higher TIT value. Also, there are more outliers in high-density conditions. This indicates that vehicles experience more severe near-collision events during Sybil attacks.

After applying the mitigation framework, the mean TIT is reduced to $1.911$, $3.694$, and $4.972$ for 30\%, 60\%, and 90\% traffic flow conditions, respectively, values that are close to those observed in the no attack condition. The framework also reduced outliers, indicating that it can effectively reduce the severity of near-collision events for many vehicles.
Similar to the t-test analysis for TET, a paired t-test analysis with a significance level of $0.005$ was performed for TIT to compare the cumulative collision risk in (i) No Attack versus Attack without mitigation, and (ii) Attack without mitigation versus Attack with mitigation conditions. For the first comparison condition, the null hypothesis is that the attack does not increase the mean TIT relative to the baseline condition, and the alternative hypothesis tests whether the attack does increase the mean TIT. For the second comparison, the null hypothesis assumes that the mitigation framework does not reduce the mean TIT from the attack without mitigation, and the alternative hypothesis tests whether mitigation reduces the mean TIT. The statistical results show extremely small p-values, ranging from ${10}^{-10}$to ${10}^{-39}$ across all traffic densities. Hence, we can reject the null hypothesis for both comparisons. These findings demonstrate that Sybil attacks substantially increased the severity of near-collision events, while the DT-based mitigation mechanism significantly reduces this effect. 

\subsubsection{Latency Analysis}

\begin{figure}
	\centering
	\includegraphics[width=\columnwidth]{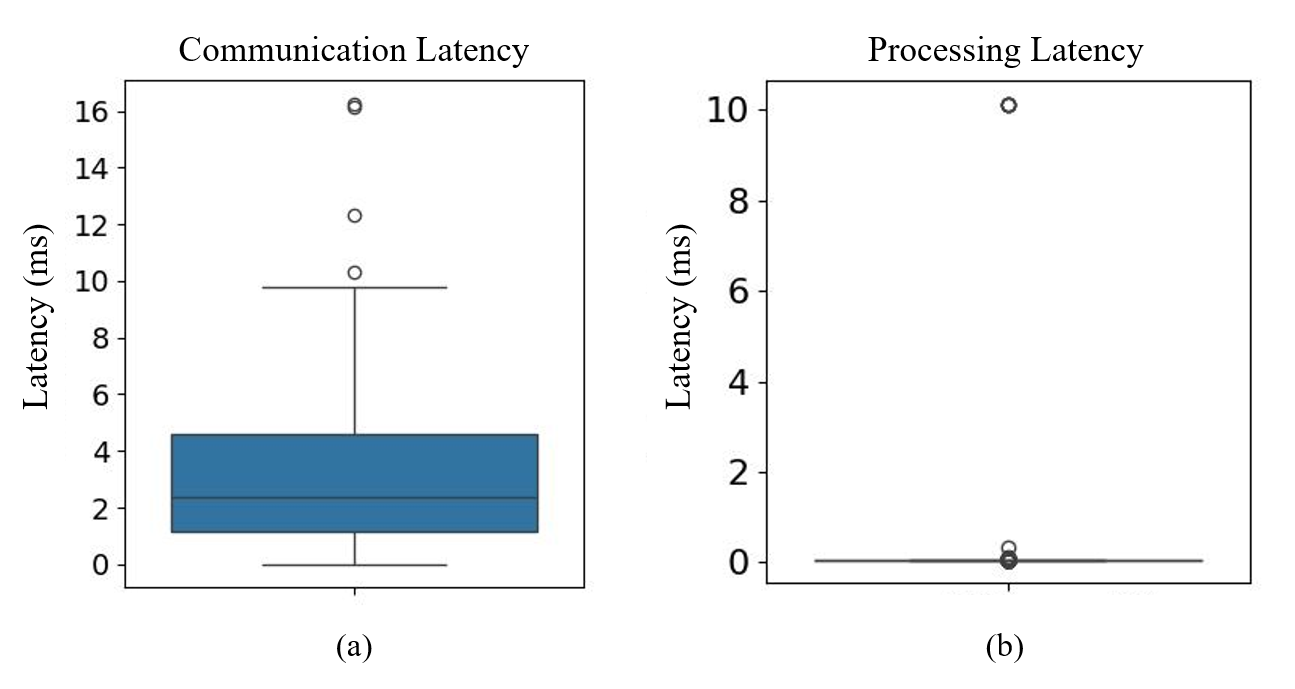}
	\caption{(a) Communication latency and (b) Processing latency of the best-performing model.}
	\label{FIG12}
\end{figure}

To evaluate the latency in detecting the fake vehicles in a Sybil attack, two components of the latency are considered: a) communication or transmission delay and b) processing delay of the model. The communication delay is the time required for BSM packets to travel from the sender vehicle to the receiver vehicle via the C-V2X radio channel. This includes both network transmission latency and potential queuing delays. The transmission latency is the time required for a BSM packet to reach the receiver radio after being sent by the transmitter. The queuing delay is the time BSM packets spend inside the receiver radio buffer. The processing latency of the Sybil attack detection model has two components: a) generating embeddings from the BSM sequence, and b) inference time. Fig. \ref{FIG12} illustrates the box plots of communication latency and processing latency of the best-performing model during the experiments.

During the field test, the maximum observed transmission delay was $16.13$ milliseconds, with a median value of $2.385$ milliseconds. The model's processing time was measured as the total time taken to generate an embedding from an input and to produce a prediction. This measurement was performed using the validation set with the best-performing model parameters. A maximum processing time of $10.124$ milliseconds was observed during the measurement, with a median value of less than $0.1$ milliseconds. In the worst-case scenario, the end-to-end latency is $26.38$ milliseconds, which is the sum of the maximum communication and processing latencies obtained from the experiments. According to the SAE J2735 standard, the maximum allowable latency for safety applications is $100$ milliseconds \cite{7}, and the latency of our framework remains well below this threshold, even in the worst-case scenario. 

\subsubsection{Computational Resource Consumption Analysis}

The DT prototype developed in this study was also evaluated for its computational resource consumption. The framework was executed on a hardware system equipped with an Intel Core i5-8250U processor (4 cores, 8 threads) and 16 GB of RAM. The simulation framework consists of the SUMO traffic simulator and a Python-based DT control and analysis script. During runtime, SUMO used approximately 40 MB of memory, while the Python script also consumed around 37 MB, totaling 77 MB of RAM, which represents less than 0.5\% of the available RAM. Simultaneously, CPU usage remained under 2\% for both these components. The model was trained and tested on a V100 GPU with 8GB of RAM. The NVIDIA V100 GPU provides high-performance computing to accelerate the inference process for the Sybil attack detection model, significantly reducing computation time. compared to conventional CPU processing. All metrics confirm that the framework operates well within the capacity of modern in-vehicle processors, such as NVIDIA Jetson AGX Xavier, Jetson AGX Orin, or similar processors, which have optimized CUDA and Tensor Cores designed for efficient AI inference \cite{46,47}.

\section{Conclusion}
 
This study presented an in-vehicle DT-based collision warning framework with Sybil attack detection capability. By combining a TCN for learning temporal patterns in vehicle trajectory data with an HNSW algorithm for efficient similarity-based classification, the framework enables reliable identification of Sybil vehicles under diverse traffic conditions. The results demonstrate that the model effectively leverages temporal context to distinguish between real and fake vehicles while maintaining stable and efficient convergence across varying traffic densities. The findings further highlight that accurate detection in dense traffic environments requires more temporal context; however, the model converges within a limited number of training epochs, suggesting efficient learning and low computational overhead. Leveraging the fake vehicle detection capability, our framework enhances the reliability of collision warning systems by mitigating the adverse effects of Sybil attacks, thereby reducing both the exposure to and severity of near-collision events in the CV collision avoidance environment. Statistical analysis results also confirmed that, across varying traffic densities, the mitigation approach is effective in reducing both the exposure and severity of unsafe events. Furthermore, the framework satisfies the latency requirements of real-time safety applications while maintaining a low computational footprint compatible with modern in-vehicle processors, making it suitable for real-world deployments.

Another key strength of our approach lies in its fully in-vehicle and decentralized design. Unlike approaches that rely on RSUs or centralized infrastructure, our framework operates independently within the vehicle, making it resilient to communication issues such as signal attenuation or loss of connectivity with RSUs. Moreover, by eliminating dependence on centralized detection mechanisms, the framework avoids the risk of single points of failure, thereby improving system robustness and reliability in safety-critical scenarios. 

Overall, our framework offers a robust, efficient, and practical solution for enhancing vehicular safety by integrating Sybil attack detection directly into the collision warning process, while ensuring resilience, scalability, and reliable operation across diverse and dynamic traffic environments.

Although the Sybil attack data were collected through field experiments, it should be noted that the experiments were conducted in a controlled environment involving only two vehicles operating on a two-lane highway. In addition, we have considered 100\% CVs on the road during the simulation. Consequently, the experimental configuration does not fully represent the variability and complexity of mixed traffic on multi-lane roads, where genuine vehicles may frequently overlap with Sybil vehicle positions. Future work will address these limitations by validating the framework under real-world multi-vehicle traffic conditions to better assess its robustness and generalizability.

\section*{Acknowledgment}
This work is based upon the work supported by the National Center for Transportation Cybersecurity and Resiliency (TraCR) (a US Department of Transportation National University Transportation Center) headquartered at Clemson University, Clemson, South Carolina, USA. Any opinions, findings, conclusions, and recommendations expressed in this material are those of the author(s) and do not necessarily reflect the views of TraCR, and the US Government assumes no liability for the contents or use thereof.

\printcredits

\section*{Funding}
This research was funded by the National Center for Transportation Cybersecurity and Resiliency (TraCR) (a US Department of Transportation National University Transportation Center) headquartered at Clemson University, Clemson, South Carolina, USA, under Grants: \\69A3552344812, 69A3552348317.

\section*{Data availability statement}
Data will be shared upon request.

\section*{Declaration of competing interest}
The authors declare that they have no conflict of interest.

\bibliographystyle{elsarticle-num}
\bibliography{cas-refs}

\begin{thebibliography}{10}
\expandafter\ifx\csname url\endcsname\relax
  \def\url#1{\texttt{#1}}\fi
\expandafter\ifx\csname urlprefix\endcsname\relax\def\urlprefix{URL }\fi
\expandafter\ifx\csname href\endcsname\relax
  \def\href#1#2{#2} \def\path#1{#1}\fi

\bibitem{1}
K.~C. Dey, L.~Yan, X.~Wang, Y.~Wang, H.~Shen, M.~Chowdhury, L.~Yu, C.~Qiu, V.~Soundararaj, \href{http://ieeexplore.ieee.org/document/7314936/}{A {Review} of {Communication}, {Driver} {Characteristics}, and {Controls} {Aspects} of {Cooperative} {Adaptive} {Cruise} {Control} ({CACC})}, IEEE Transactions on Intelligent Transportation Systems 17~(2) (2016) 491--509.
\newblock \href {https://doi.org/10.1109/TITS.2015.2483063} {\path{doi:10.1109/TITS.2015.2483063}}.
\newline\urlprefix\url{http://ieeexplore.ieee.org/document/7314936/}

\bibitem{2}
K.~Abboud, H.~A. Omar, W.~Zhuang, \href{http://ieeexplore.ieee.org/document/7513432/}{Interworking of {DSRC} and {Cellular} {Network} {Technologies} for {V2X} {Communications}: {A} {Survey}}, IEEE Transactions on Vehicular Technology 65~(12) (2016) 9457--9470.
\newblock \href {https://doi.org/10.1109/TVT.2016.2591558} {\path{doi:10.1109/TVT.2016.2591558}}.
\newline\urlprefix\url{http://ieeexplore.ieee.org/document/7513432/}

\bibitem{3}
Z.~Huang, S.~Chen, Y.~Pian, Z.~Sheng, S.~Ahn, D.~A. Noyce, \href{https://ieeexplore.ieee.org/document/10556814/}{Toward {C}-{V2X} {Enabled} {Connected} {Transportation} {System}: {RSU}-{Based} {Cooperative} {Localization} {Framework} for {Autonomous} {Vehicles}}, IEEE Transactions on Intelligent Transportation Systems 25~(10) (2024) 13417--13431.
\newblock \href {https://doi.org/10.1109/TITS.2024.3410185} {\path{doi:10.1109/TITS.2024.3410185}}.
\newline\urlprefix\url{https://ieeexplore.ieee.org/document/10556814/}

\bibitem{4}
J.~Petit, S.~E. Shladover, \href{https://ieeexplore.ieee.org/document/6899663}{Potential {Cyberattacks} on {Automated} {Vehicles}}, IEEE Transactions on Intelligent Transportation Systems (2014) 1--11\href {https://doi.org/10.1109/TITS.2014.2342271} {\path{doi:10.1109/TITS.2014.2342271}}.
\newline\urlprefix\url{https://ieeexplore.ieee.org/document/6899663}

\bibitem{5}
S.~Checkoway, D.~McCoy, B.~Kantor, D.~Anderson, H.~Shacham, S.~Savage, K.~Koscher, A.~Czeskis, F.~Roesner, T.~Kohno, \href{https://www.usenix.org/conference/usenix-security-11/comprehensive-experimental-analyses-automotive-attack-surfaces}{Comprehensive {Experimental} {Analyses} of {Automotive} {Attack} {Surfaces}}, in: 20th {USENIX} {Security} {Symposium} ({USENIX} {Security} 11), USENIX Association, San Francisco, CA, 2011.
\newline\urlprefix\url{https://www.usenix.org/conference/usenix-security-11/comprehensive-experimental-analyses-automotive-attack-surfaces}

\bibitem{6}
C.~H. O.~O. Quevedo, A.~M. B.~C. Quevedo, G.~A. Campos, R.~L. Gomes, J.~Celestino, A.~Serhrouchni, \href{https://ieeexplore.ieee.org/document/9149371/}{An {Intelligent} {Mechanism} for {Sybil} {Attacks} {Detection} in {VANETs}}, in: {ICC} 2020 - 2020 {IEEE} {International} {Conference} on {Communications} ({ICC}), IEEE, Dublin, Ireland, 2020, pp. 1--6.
\newblock \href {https://doi.org/10.1109/ICC40277.2020.9149371} {\path{doi:10.1109/ICC40277.2020.9149371}}.
\newline\urlprefix\url{https://ieeexplore.ieee.org/document/9149371/}

\bibitem{7}
Y.~Zhong, B.~Yang, Y.~Li, H.~Yang, X.~Li, Y.~Zhang, \href{https://ieeexplore.ieee.org/document/10387880/}{Tackling {Sybil} {Attacks} in {Intelligent} connected vehicles: {A} {Review} of {Machine} {Learning} and {Deep} {Learning} {Techniques}}, in: 2023 8th {International} {Conference} on {Computational} {Intelligence} and {Applications} ({ICCIA}), IEEE, Haikou, China, 2023, pp. 8--12.
\newblock \href {https://doi.org/10.1109/ICCIA59741.2023.00010} {\path{doi:10.1109/ICCIA59741.2023.00010}}.
\newline\urlprefix\url{https://ieeexplore.ieee.org/document/10387880/}

\bibitem{8}
W.~Whyte, A.~Weimerskirch, V.~Kumar, T.~Hehn, \href{http://ieeexplore.ieee.org/document/6737583/}{A security credential management system for {V2V} communications}, in: 2013 {IEEE} {Vehicular} {Networking} {Conference}, IEEE, Boston, MA, USA, 2013, pp. 1--8.
\newblock \href {https://doi.org/10.1109/VNC.2013.6737583} {\path{doi:10.1109/VNC.2013.6737583}}.
\newline\urlprefix\url{http://ieeexplore.ieee.org/document/6737583/}

\bibitem{9}
B.~Hammi, Y.~M. Idir, S.~Zeadally, R.~Khatoun, J.~Nebhen, \href{https://ieeexplore.ieee.org/document/9757751/}{Is it {Really} {Easy} to {Detect} {Sybil} {Attacks} in {C}-{ITS} {Environments}: {A} {Position} {Paper}}, IEEE Transactions on Intelligent Transportation Systems 23~(10) (2022) 18273--18287.
\newblock \href {https://doi.org/10.1109/TITS.2022.3165513} {\path{doi:10.1109/TITS.2022.3165513}}.
\newline\urlprefix\url{https://ieeexplore.ieee.org/document/9757751/}

\bibitem{10}
Z.~Yang, K.~Zhang, L.~Lei, K.~Zheng, \href{https://ieeexplore.ieee.org/document/8474346/}{A {Novel} {Classifier} {Exploiting} {Mobility} {Behaviors} for {Sybil} {Detection} in {Connected} {Vehicle} {Systems}}, IEEE Internet of Things Journal 6~(2) (2019) 2626--2636.
\newblock \href {https://doi.org/10.1109/JIOT.2018.2872456} {\path{doi:10.1109/JIOT.2018.2872456}}.
\newline\urlprefix\url{https://ieeexplore.ieee.org/document/8474346/}

\bibitem{11}
A.~Muhamad, M.~Elhadef, \href{http://link.springer.com/10.1007/978-981-13-1328-8_71}{Sybil {Attacks} in {Intelligent} {Vehicular} {Ad} {Hoc} {Networks}: {A} {Review}}, in: J.~J. Park, V.~Loia, K.-K.~R. Choo, G.~Yi (Eds.), Advanced {Multimedia} and {Ubiquitous} {Engineering}, Vol. 518, Springer Singapore, Singapore, 2019, pp. 547--555, series Title: Lecture Notes in Electrical Engineering.
\newblock \href {https://doi.org/10.1007/978-981-13-1328-8_71} {\path{doi:10.1007/978-981-13-1328-8_71}}.
\newline\urlprefix\url{http://link.springer.com/10.1007/978-981-13-1328-8_71}

\bibitem{12}
S.~E. Carpenter, M.~L. Sichitiu, \href{https://dl.acm.org/doi/10.1145/2756509.2756512}{An obstacle model implementation for evaluating radio shadowing with ns-3}, in: Proceedings of the 2015 {Workshop} on ns-3, ACM, Barcelona Spain, 2015, pp. 17--24.
\newblock \href {https://doi.org/10.1145/2756509.2756512} {\path{doi:10.1145/2756509.2756512}}.
\newline\urlprefix\url{https://dl.acm.org/doi/10.1145/2756509.2756512}

\bibitem{13}
L.~Li, S.~Aslam, A.~Wileman, S.~Perinpanayagam, \href{https://ieeexplore.ieee.org/document/9656111/}{Digital {Twin} in {Aerospace} {Industry}: {A} {Gentle} {Introduction}}, IEEE Access 10 (2022) 9543--9562.
\newblock \href {https://doi.org/10.1109/ACCESS.2021.3136458} {\path{doi:10.1109/ACCESS.2021.3136458}}.
\newline\urlprefix\url{https://ieeexplore.ieee.org/document/9656111/}

\bibitem{14}
W.~A. Ali, M.~Roccotelli, M.~P. Fanti, \href{https://ieeexplore.ieee.org/document/9804017/}{Digital {Twin} in {Intelligent} {Transportation} {Systems}: a {Review}}, in: 2022 8th {International} {Conference} on {Control}, {Decision} and {Information} {Technologies} ({CoDIT}), IEEE, Istanbul, Turkey, 2022, pp. 576--581.
\newblock \href {https://doi.org/10.1109/CoDIT55151.2022.9804017} {\path{doi:10.1109/CoDIT55151.2022.9804017}}.
\newline\urlprefix\url{https://ieeexplore.ieee.org/document/9804017/}

\bibitem{15}
L.~Bao, Q.~Wang, Y.~Jiang, \href{https://ieeexplore.ieee.org/document/9666030/}{Review of {Digital} twin for intelligent transportation system}, in: 2021 {International} {Conference} on {Information} {Control}, {Electrical} {Engineering} and {Rail} {Transit} ({ICEERT}), IEEE, Lanzhou, China, 2021, pp. 309--315.
\newblock \href {https://doi.org/10.1109/ICEERT53919.2021.00064} {\path{doi:10.1109/ICEERT53919.2021.00064}}.
\newline\urlprefix\url{https://ieeexplore.ieee.org/document/9666030/}

\bibitem{16}
X.~Gu, M.~Abdel-Aty, Q.~Xiang, Q.~Cai, J.~Yuan, \href{https://linkinghub.elsevier.com/retrieve/pii/S0001457518309631}{Utilizing {UAV} video data for in-depth analysis of drivers’ crash risk at interchange merging areas}, Accident Analysis \& Prevention 123 (2019) 159--169.
\newblock \href {https://doi.org/10.1016/j.aap.2018.11.010} {\path{doi:10.1016/j.aap.2018.11.010}}.
\newline\urlprefix\url{https://linkinghub.elsevier.com/retrieve/pii/S0001457518309631}

\bibitem{17}
X.~Gu, Q.~Cai, J.~Lee, Q.~Xiang, Y.~Ma, X.~Xu, \href{https://www.tandfonline.com/doi/full/10.1080/15389588.2020.1734581}{Proactive crash risk prediction modeling for merging assistance system at interchange merging areas}, Traffic Injury Prevention 21~(3) (2020) 234--240.
\newblock \href {https://doi.org/10.1080/15389588.2020.1734581} {\path{doi:10.1080/15389588.2020.1734581}}.
\newline\urlprefix\url{https://www.tandfonline.com/doi/full/10.1080/15389588.2020.1734581}

\bibitem{18}
S.~Li, Q.~Xiang, Y.~Ma, X.~Gu, H.~Li, \href{https://www.mdpi.com/1660-4601/13/11/1157}{Crash {Risk} {Prediction} {Modeling} {Based} on the {Traffic} {Conflict} {Technique} and a {Microscopic} {Simulation} for {Freeway} {Interchange} {Merging} {Areas}}, International Journal of Environmental Research and Public Health 13~(11) (2016) 1157.
\newblock \href {https://doi.org/10.3390/ijerph13111157} {\path{doi:10.3390/ijerph13111157}}.
\newline\urlprefix\url{https://www.mdpi.com/1660-4601/13/11/1157}

\bibitem{19}
Y.~Zheng, S.~Yang, H.~Cheng, \href{http://link.springer.com/10.1007/s12652-018-0911-3}{An application framework of digital twin and its case study}, Journal of Ambient Intelligence and Humanized Computing 10~(3) (2019) 1141--1153.
\newblock \href {https://doi.org/10.1007/s12652-018-0911-3} {\path{doi:10.1007/s12652-018-0911-3}}.
\newline\urlprefix\url{http://link.springer.com/10.1007/s12652-018-0911-3}

\bibitem{20}
M.~Masi, G.~P. Sellitto, H.~Aranha, T.~Pavleska, \href{https://link.springer.com/10.1007/s10270-022-01075-0}{Securing critical infrastructures with a cybersecurity digital twin}, Software and Systems Modeling 22~(2) (2023) 689--707.
\newblock \href {https://doi.org/10.1007/s10270-022-01075-0} {\path{doi:10.1007/s10270-022-01075-0}}.
\newline\urlprefix\url{https://link.springer.com/10.1007/s10270-022-01075-0}

\bibitem{21}
M.~Korman, M.~Välja, G.~Björkman, M.~Ekstedt, A.~Vernotte, R.~Lagerström, \href{https://dl.acm.org/doi/10.1145/3055386.3055393}{Analyzing the {Effectiveness} of {Attack} {Countermeasures} in a {SCADA} {System}}, in: Proceedings of the 2nd {Workshop} on {Cyber}-{Physical} {Security} and {Resilience} in {Smart} {Grids}, ACM, Pittsburgh PA USA, 2017, pp. 73--78.
\newblock \href {https://doi.org/10.1145/3055386.3055393} {\path{doi:10.1145/3055386.3055393}}.
\newline\urlprefix\url{https://dl.acm.org/doi/10.1145/3055386.3055393}

\bibitem{22}
M.~Ali, G.~Kaddoum, W.-T. Li, C.~Yuen, M.~Tariq, H.~V. Poor, \href{https://ieeexplore.ieee.org/document/10220152/}{A {Smart} {Digital} {Twin} {Enabled} {Security} {Framework} for {Vehicle}-to-{Grid} {Cyber}-{Physical} {Systems}}, IEEE Transactions on Information Forensics and Security 18 (2023) 5258--5271.
\newblock \href {https://doi.org/10.1109/TIFS.2023.3305916} {\path{doi:10.1109/TIFS.2023.3305916}}.
\newline\urlprefix\url{https://ieeexplore.ieee.org/document/10220152/}

\bibitem{23}
M.~Eckhart, A.~Ekelhart, \href{https://dl.acm.org/doi/10.1145/3264888.3264892}{A {Specification}-based {State} {Replication} {Approach} for {Digital} {Twins}}, in: Proceedings of the 2018 {Workshop} on {Cyber}-{Physical} {Systems} {Security} and {PrivaCy}, ACM, Toronto Canada, 2018, pp. 36--47.
\newblock \href {https://doi.org/10.1145/3264888.3264892} {\path{doi:10.1145/3264888.3264892}}.
\newline\urlprefix\url{https://dl.acm.org/doi/10.1145/3264888.3264892}

\bibitem{24}
M.~Eckhart, A.~Ekelhart, E.~Weippl, \href{https://ieeexplore.ieee.org/document/8869197/}{Enhancing {Cyber} {Situational} {Awareness} for {Cyber}-{Physical} {Systems} through {Digital} {Twins}}, in: 2019 24th {IEEE} {International} {Conference} on {Emerging} {Technologies} and {Factory} {Automation} ({ETFA}), IEEE, Zaragoza, Spain, 2019, pp. 1222--1225.
\newblock \href {https://doi.org/10.1109/ETFA.2019.8869197} {\path{doi:10.1109/ETFA.2019.8869197}}.
\newline\urlprefix\url{https://ieeexplore.ieee.org/document/8869197/}

\bibitem{25}
A.~Kummerow, D.~Rosch, C.~Monsalve, S.~Nicolai, P.~Bretschneider, C.~Brosinsky, D.~Westermann, \href{https://ieeexplore.ieee.org/document/8810711/}{Challenges and opportunities for phasor data based event detection in transmission control centers under cyber security constraints}, in: 2019 {IEEE} {Milan} {PowerTech}, IEEE, Milan, Italy, 2019, pp. 1--6.
\newblock \href {https://doi.org/10.1109/PTC.2019.8810711} {\path{doi:10.1109/PTC.2019.8810711}}.
\newline\urlprefix\url{https://ieeexplore.ieee.org/document/8810711/}

\bibitem{26}
J.~Guo, X.~Wu, H.~Liang, J.~Hu, B.~Liu, \href{https://ieeexplore.ieee.org/document/9270257/}{Digital-twin based {Power} {Supply} {System} {Modeling} and {Analysis} for {Urban} {Rail} {Transportation}}, in: 2020 {IEEE} {International} {Conference} on {Energy} {Internet} ({ICEI}), IEEE, Sydney, NSW, Australia, 2020, pp. 74--79.
\newblock \href {https://doi.org/10.1109/ICEI49372.2020.00022} {\path{doi:10.1109/ICEI49372.2020.00022}}.
\newline\urlprefix\url{https://ieeexplore.ieee.org/document/9270257/}

\bibitem{27}
S.~Wang, F.~Zhang, T.~Qin, \href{https://iopscience.iop.org/article/10.1088/1742-6596/1802/4/042045}{Research on the {Construction} of {Highway} {Traffic} {Digital} {Twin} {System} {Based} on {3D} {GIS} {Technology}}, Journal of Physics: Conference Series 1802~(4) (2021) 042045.
\newblock \href {https://doi.org/10.1088/1742-6596/1802/4/042045} {\path{doi:10.1088/1742-6596/1802/4/042045}}.
\newline\urlprefix\url{https://iopscience.iop.org/article/10.1088/1742-6596/1802/4/042045}

\bibitem{28}
J.-S. Kang, K.~Chung, E.~J. Hong, \href{https://link.springer.com/10.1007/s11042-021-10649-x}{Multimedia knowledge‐based bridge health monitoring using digital twin}, Multimedia Tools and Applications 80~(26-27) (2021) 34609--34624.
\newblock \href {https://doi.org/10.1007/s11042-021-10649-x} {\path{doi:10.1007/s11042-021-10649-x}}.
\newline\urlprefix\url{https://link.springer.com/10.1007/s11042-021-10649-x}

\bibitem{29}
T.~Li, Z.~Bian, H.~Lei, F.~Zuo, Y.-T. Yang, Q.~Zhu, Z.~Li, Z.~Chen, K.~Ozbay, \href{https://arxiv.org/abs/2407.15025}{Digital {Twin}-based {Driver} {Risk}-{Aware} {Intelligent} {Mobility} {Analytics} for {Urban} {Transportation} {Management}}, version Number: 1 (2024).
\newblock \href {https://doi.org/10.48550/ARXIV.2407.15025} {\path{doi:10.48550/ARXIV.2407.15025}}.
\newline\urlprefix\url{https://arxiv.org/abs/2407.15025}

\bibitem{30}
B.~Mokhtar, M.~Azab, \href{https://linkinghub.elsevier.com/retrieve/pii/S1110016815001246}{Survey on {Security} {Issues} in {Vehicular} {Ad} {Hoc} {Networks}}, Alexandria Engineering Journal 54~(4) (2015) 1115--1126.
\newblock \href {https://doi.org/10.1016/j.aej.2015.07.011} {\path{doi:10.1016/j.aej.2015.07.011}}.
\newline\urlprefix\url{https://linkinghub.elsevier.com/retrieve/pii/S1110016815001246}

\bibitem{31}
M.~Asad, S.~Otoum, \href{https://ieeexplore.ieee.org/document/11161123/}{{FL}-{SATS}: {Federated} {Learning} for {Sybil} {Attack} {Detection} in {Transportation} {System}}, in: {ICC} 2025 - {IEEE} {International} {Conference} on {Communications}, IEEE, Montreal, QC, Canada, 2025, pp. 3376--3381.
\newblock \href {https://doi.org/10.1109/ICC52391.2025.11161123} {\path{doi:10.1109/ICC52391.2025.11161123}}.
\newline\urlprefix\url{https://ieeexplore.ieee.org/document/11161123/}

\bibitem{32}
S.~Azam, M.~Bibi, R.~Riaz, S.~S. Rizvi, S.~J. Kwon, \href{https://www.mdpi.com/1424-8220/22/18/6934}{Collaborative {Learning} {Based} {Sybil} {Attack} {Detection} in {Vehicular} {AD}-{HOC} {Networks} ({VANETS})}, Sensors 22~(18) (2022) 6934.
\newblock \href {https://doi.org/10.3390/s22186934} {\path{doi:10.3390/s22186934}}.
\newline\urlprefix\url{https://www.mdpi.com/1424-8220/22/18/6934}

\bibitem{33}
Y.~Yao, B.~Xiao, G.~Wu, X.~Liu, Z.~Yu, K.~Zhang, X.~Zhou, \href{https://ieeexplore.ieee.org/document/8356112/}{Multi-{Channel} {Based} {Sybil} {Attack} {Detection} in {Vehicular} {Ad} {Hoc} {Networks} {Using} {RSSI}}, IEEE Transactions on Mobile Computing 18~(2) (2019) 362--375.
\newblock \href {https://doi.org/10.1109/TMC.2018.2833849} {\path{doi:10.1109/TMC.2018.2833849}}.
\newline\urlprefix\url{https://ieeexplore.ieee.org/document/8356112/}

\bibitem{34}
S.~Rakhi, K.~R. Shobha, \href{https://ieeexplore.ieee.org/document/10179918/}{{LCSS} {Based} {Sybil} {Attack} {Detection} and {Avoidance} in {Clustered} {Vehicular} {Networks}}, IEEE Access 11 (2023) 75179--75190.
\newblock \href {https://doi.org/10.1109/ACCESS.2023.3294469} {\path{doi:10.1109/ACCESS.2023.3294469}}.
\newline\urlprefix\url{https://ieeexplore.ieee.org/document/10179918/}

\bibitem{35}
A.~El~Attar, M.~Ali~Awali, R.~Khatoun, M.~Hatoum, K.~Samrouth, \href{https://ieeexplore.ieee.org/document/10911022/}{Detecting {Malicious} {Artificial} {Congestion} in {Connected} {Cars} {Environment}}, in: 2025 5th {IEEE} {Middle} {East} and {North} {Africa} {Communications} {Conference} ({MENACOMM}), IEEE, Byblos, Lebanon, 2025, pp. 1--8.
\newblock \href {https://doi.org/10.1109/MENACOMM62946.2025.10911022} {\path{doi:10.1109/MENACOMM62946.2025.10911022}}.
\newline\urlprefix\url{https://ieeexplore.ieee.org/document/10911022/}

\bibitem{36}
M.~Baza, M.~Nabil, M.~M. E.~A. Mahmoud, N.~Bewermeier, K.~Fidan, W.~Alasmary, M.~Abdallah, \href{https://ieeexplore.ieee.org/document/9091099/}{Detecting {Sybil} {Attacks} {Using} {Proofs} of {Work} and {Location} in {VANETs}}, IEEE Transactions on Dependable and Secure Computing 19~(1) (2022) 39--53.
\newblock \href {https://doi.org/10.1109/TDSC.2020.2993769} {\path{doi:10.1109/TDSC.2020.2993769}}.
\newline\urlprefix\url{https://ieeexplore.ieee.org/document/9091099/}

\bibitem{37}
T.~Guven, Z.~C. Taysi, \href{https://link.springer.com/10.1007/s12083-025-02058-w}{Creating a realistic sybil attack dataset for inter-vehicle communication}, Peer-to-Peer Networking and Applications 18~(4) (2025) 234.
\newblock \href {https://doi.org/10.1007/s12083-025-02058-w} {\path{doi:10.1007/s12083-025-02058-w}}.
\newline\urlprefix\url{https://link.springer.com/10.1007/s12083-025-02058-w}

\bibitem{38}
Y.~Chen, Y.~Lai, Z.~Zhang, H.~Li, Y.~Wang, \href{https://ieeexplore.ieee.org/document/9829793/}{Malicious attack detection based on traffic-flow information fusion}, in: 2022 {IFIP} {Networking} {Conference} ({IFIP} {Networking}), IEEE, Catania, Italy, 2022, pp. 1--9.
\newblock \href {https://doi.org/10.23919/IFIPNetworking55013.2022.9829793} {\path{doi:10.23919/IFIPNetworking55013.2022.9829793}}.
\newline\urlprefix\url{https://ieeexplore.ieee.org/document/9829793/}

\bibitem{39}
A.~Enan, A.~A. Mamun, J.~M. Tine, J.~Mwakalonge, D.~A. Indah, G.~Comert, M.~Chowdhury, \href{https://dl.acm.org/doi/10.1145/3643823}{Basic {Safety} {Message} {Generation} through a {Video}-based {Analytics} for {Potential} {Safety} {Applications}}, ACM Journal on Autonomous Transportation Systems 1~(4) (2024) 1--26.
\newblock \href {https://doi.org/10.1145/3643823} {\path{doi:10.1145/3643823}}.
\newline\urlprefix\url{https://dl.acm.org/doi/10.1145/3643823}

\bibitem{40}
{V2X Core Technical Committee}, \href{https://saemobilus.sae.org/standards/j2735_202409-v2x-communications-message-set-dictionary}{{V2X} {Communications} {Message} {Set} {Dictionary}}.
\newblock \href {https://doi.org/10.4271/J2735_202409} {\path{doi:10.4271/J2735_202409}}.
\newline\urlprefix\url{https://saemobilus.sae.org/standards/j2735_202409-v2x-communications-message-set-dictionary}

\bibitem{41}
S.~Bai, J.~Z. Kolter, V.~Koltun, \href{https://arxiv.org/abs/1803.01271}{An {Empirical} {Evaluation} of {Generic} {Convolutional} and {Recurrent} {Networks} for {Sequence} {Modeling}}, version Number: 2 (2018).
\newblock \href {https://doi.org/10.48550/ARXIV.1803.01271} {\path{doi:10.48550/ARXIV.1803.01271}}.
\newline\urlprefix\url{https://arxiv.org/abs/1803.01271}

\bibitem{42}
Y.~A. Malkov, D.~A. Yashunin, \href{https://ieeexplore.ieee.org/document/8594636/}{Efficient and {Robust} {Approximate} {Nearest} {Neighbor} {Search} {Using} {Hierarchical} {Navigable} {Small} {World} {Graphs}}, IEEE Transactions on Pattern Analysis and Machine Intelligence 42~(4) (2020) 824--836.
\newblock \href {https://doi.org/10.1109/TPAMI.2018.2889473} {\path{doi:10.1109/TPAMI.2018.2889473}}.
\newline\urlprefix\url{https://ieeexplore.ieee.org/document/8594636/}

\bibitem{43}
E.~Williams, Aviation {Formulary} v1. 47 (2013), URL: https://www. edwilliams. org/avform. htm.

\bibitem{44}
\href{https://www.trb.org/Main/Blurbs/164718.aspx}{Highway {Capacity} {Manual} 2010 ({HCM2010})}.
\newline\urlprefix\url{https://www.trb.org/Main/Blurbs/164718.aspx}

\bibitem{45}
M.~S. Rahman, M.~Abdel-Aty, J.~Lee, M.~H. Rahman, \href{https://linkinghub.elsevier.com/retrieve/pii/S0968090X18310349}{Safety benefits of arterials’ crash risk under connected and automated vehicles}, Transportation Research Part C: Emerging Technologies 100 (2019) 354--371.
\newblock \href {https://doi.org/10.1016/j.trc.2019.01.029} {\path{doi:10.1016/j.trc.2019.01.029}}.
\newline\urlprefix\url{https://linkinghub.elsevier.com/retrieve/pii/S0968090X18310349}

\bibitem{46}
\href{https://www.nvidia.com/en-us/autonomous-machines/embedded-systems/jetson-agx-xavier/}{Deploy {AI}-{Powered} {Autonomous} {Machines} at {Scale}}.
\newline\urlprefix\url{https://www.nvidia.com/en-us/autonomous-machines/embedded-systems/jetson-agx-xavier/}

\bibitem{47}
\href{https://www.nvidia.com/en-us/autonomous-machines/embedded-systems/jetson-orin/}{{NVIDIA} {Jetson} {AGX} {Orin}}.
\newline\urlprefix\url{https://www.nvidia.com/en-us/autonomous-machines/embedded-systems/jetson-orin/}

\end{thebibliography}


\end{document}